\documentclass[english]{article}
\usepackage[margin=1.0in]{geometry}
\usepackage[T1]{fontenc}
\usepackage[latin1]{inputenc}
\usepackage{subfig}
\usepackage{float}
\usepackage{graphicx}
\graphicspath{{img/}}
\usepackage{setspace}
\usepackage{amssymb}
\usepackage{babel}
\usepackage{acronym}
\usepackage{units}
\usepackage{natbib}
\usepackage{url}
\usepackage{acronym}
\usepackage{graphicx}
\usepackage{amsmath, amsthm, amssymb, amsfonts, mathtools}
\usepackage{pbox}
\usepackage{placeins}

\renewcommand{\refeq}[1]{{Eq.~(\ref{#1})}}

\newcommand{\reffig}[1]{{Fig.~\ref{#1}}}

\newcommand{\refsec}[1]{{Sec.~\ref{#1}}}

\renewcommand{\cite}{\citep}

\usepackage{todonotes}

\makeatother
\begin{document}

\acrodef{AC}[AC]{Arrenhius \& Current}
\acrodef{AER}[AER]{Address Event Representation}
\acrodef{AEX}[AEX]{AER EXtension board}
\acrodef{AMDA}[AMDA]{``AER Motherboard with D/A converters''}
\acrodef{API}[API]{Application Programming Interface}
\acrodef{BM}[BM]{Boltzmann Machine}
\acrodef{CAVIAR}[CAVIAR]{Convolution AER Vision Architecture for Real-Time}
\acrodef{CCN}[CCN]{Cooperative and Competitive Network}
\acrodef{CD}[CD]{Contrastive Divergence}
\acrodef{CMOS}[CMOS]{Complementary Metal--Oxide--Semiconductor}
\acrodef{COTS}[COTS]{Commercial Off-The-Shelf}
\acrodef{CPU}[CPU]{Central Processing Unit}
\acrodef{CV}[CV]{Coefficient of Variation}
\acrodef{CV}[CV]{Coefficient of Variation}
\acrodef{DAC}[DAC]{Digital--to--Analog}
\acrodef{DBN}[DBN]{Deep Belief Network}
\acrodef{DFA}[DFA]{Deterministic Finite Automaton}
\acrodef{DFA}[DFA]{Deterministic Finite Automaton}
\acrodef{divmod3}[DIVMOD3]{divisibility of a number by 3}
\acrodef{DPE}[DPE]{Dynamic Parameter Estimation}
\acrodef{DPI}[DPI]{Differential-Pair Integrator}
\acrodef{DSP}[DSP]{Digital Signal Processor}
\acrodef{DVS}[DVS]{Dynamic Vision Sensor}
\acrodef{EDVAC}[EDVAC]{Electronic Discrete Variable Automatic Computer}
\acrodef{EIF}[EI\&F]{Exponential Integrate \& Fire}
\acrodef{EIN}[EIN]{Excitatory--Inhibitory Network}
\acrodef{EPSC}[EPSC]{Excitatory Post-Synaptic Current}
\acrodef{eRBP}[eRBP]{event-driven Random Back-Propagation}
\acrodef{EPSP}[EPSP]{Excitatory Post--Synaptic Potential}
\acrodef{FPGA}[FPGA]{Field Programmable Gate Array}
\acrodef{FSM}[FSM]{Finite State Machine}
\acrodef{GPU}[GPU]{Graphical Processing Unit}
\acrodef{HAL}[HAL]{Hardware Abstraction Layer}
\acrodef{HH}[H\&H]{Hodgkin \& Huxley}
\acrodef{HMM}[HMM]{Hidden Markov Model}
\acrodef{HW}[HW]{Hardware}
\acrodef{hWTA}[hWTA]{Hard Winner--Take--All}
\acrodef{IF2DWTA}[IF2DWTA]{Integrate \& Fire 2--Dimensional WTA}
\acrodef{IF}[I\&F]{Integrate \& Fire}
\acrodef{IFSLWTA}[IFSLWTA]{Integrate \& Fire Stop Learning WTA}
\acrodef{INCF}[INCF]{International Neuroinformatics Coordinating Facility}
\acrodef{INI}[INI]{Institute of Neuroinformatics}
\acrodef{IO}[IO]{Input-Output}
\acrodef{IPSC}[IPSC]{Inhibitory Post-Synaptic Current}
\acrodef{ISI}[ISI]{Inter--Spike Interval}
\acrodef{JFLAP}[JFLAP]{Java - Formal Languages and Automata Package}
\acrodef{LIF}[LI\&F]{Leaky Integrate \& Fire}
\acrodef{LSM}[LSM]{Liquid State Machine}
\acrodef{LTD}[LTD]{Long-Term Depression}
\acrodef{LTI}[LTI]{Linear Time-Invariant}
\acrodef{LTP}[LTP]{Long-Term Potentiation}
\acrodef{LTU}[LTU]{Linear Threshold Unit}
\acrodef{MCMC}{Markov Chain Monte Carlo}
\acrodef{NHML}[NHML]{Neuromorphic Hardware Mark-up Language}
\acrodef{NMDA}[NMDA]{NMDA}
\acrodef{NME}[NE]{Neuromorphic Engineering}
\acrodef{PCB}[PCB]{Printed Circuit Board}
\acrodef{PRC}[PRC]{Phase Response Curve}
\acrodef{PSC}[PSC]{Post-Synaptic Current}
\acrodef{PSP}[PSP]{Post--Synaptic Potential}
\acrodef{RI}[KL]{Kullback-Leibler}
\acrodef{RNN}[RNN]{Recurrent Neural Network}
\acrodef{RRAM}[RRAM]{Resistive Random-Access Memory}
\acrodef{RBM}[RBM]{Restricted Boltzmann Machine}
\acrodef{ROC}[ROC]{Receiver Operator Characteristic}
\acrodef{SAC}[SAC]{Selective Attention Chip}
\acrodef{SCD}[SCD]{Spike-Based Contrastive Divergence}
\acrodef{SCX}[SCX]{Silicon CorteX}
\acrodef{SSM}[SSM]{Synaptic Sampling Machines}
\acrodef{SNN}[SNN]{Spiking Neural Network}
\acrodef{STDP}[STDP]{Spike Time Dependent Plasticity}
\acrodef{SW}[SW]{Software}
\acrodef{sWTA}[SWTA]{Soft Winner--Take--All}
\acrodef{VHDL}[VHDL]{VHSIC Hardware Description Language}
\acrodef{VLSI}[VLSI]{Very  Large  Scale  Integration}
\acrodef{WTA}[WTA]{Winner--Take--All}
\acrodef{XML}[XML]{eXtensible Mark-up Language}
 
\title{Spiking Neural Networks for Inference and Learning: A Memristor-based Design Perspective}
\author{M. E. Fouda, F. Kurdahi, A. Eltawil, E. Neftci}
\maketitle

\begin{abstract}

On metrics of density and power efficiency, neuromorphic technologies have the potential to surpass mainstream computing technologies in tasks where real-time functionality, adaptability, and autonomy are essential. While algorithmic advances in neuromorphic computing are proceeding successfully, the potential of memristors to improve neuromorphic computing have not yet born fruit, primarily because they are often used as a drop-in replacement to conventional memory. 
However, interdisciplinary approaches anchored in machine learning theory suggest that multifactor plasticity rules matching neural and synaptic dynamics to the device capabilities can take better advantage of memristor dynamics and its stochasticity. Furthermore, such plasticity rules generally show much higher performance than that of classical \ac{STDP} rules. This chapter reviews the recent development in learning with spiking neural network models and their possible implementation with memristor-based hardware.
\end{abstract}


\section{Introduction}
Machine Learning and particularly deep learning have become the \emph{de facto} choice in solving a wide range of problems when adequate data is available. 
So far, machine learning has been mainly concerned more by the ``learning'' rather than the ``machine''. 
This focus is natural given that von Neumann computers and GPUs capable of general-purpose processing offer excellent performance per unit of monetary cost.
As the scalability of such processors hit difficult scalability and energy efficiency challenges, interest in dedicated, multicore and multiprocessor systems is increasing.
This calls for increased efforts on improving the physical instantiations of ``machines'' for machine learning. 
Physical instantiation of computations are challenging because the structure and nature of the physical substrate severely restrict the basic computational operations it can carry out. 
However, if the computations can be formulated in a way that they exploit the physics of the devices, then the efficiency and scalability can be drastically improved.
In this line of thought, the field of neuromorphic engineering is arguably the one that has attracted the most attention and effort. The field's core ideas, communicated by R. Feynmann, C. Mead and other researchers in a series of lectures called physics of computation, elaborate on the analogies between the physics of ionic channels in the brain and those of CMOS transistors \cite{Mead90_neurelec}.
By building synapses, neurons, and circuits modeled after the brain and driven by similar physical laws, neuromorphic engineers would ``understand by building'' and help engineering novel computing technologies equipped with the robustness and efficiency of the brain.
In the last decade, there have been enormous advances in building and scaling neuromorphic hardware using mixed-signal \cite{Benjamin_etal14_neurmixe,Chicca_etal13_neurelec,Park_etal14_65k-73-m,Schemmel_etal10_wafeneur} and digital \cite{Merolla_etal14_millspik,Davies_etal18_loihneur,Furber_etal14_spinproj} technologies, including embedded learning capabilities and scales achieving 1M neurons per chip.
A major limitation in these technologies is the memory required to store the state and the parameters of the system. 
For example, in both mixed-signal and digital technologies, synaptic weights are typically stored in SRAM, the densest, fastest and most energy-efficient memory we can currently locate next to the computing substrate \cite{Qiao_etal15_recoon-l,Merolla_etal14_millspik,Davies_etal18_loihneur}.
Unfortunately, SRAMs are expensive in terms of area and energy, making on-chip memory small given the computational requirements. 
In fact, learning often requires higher precision parameters to average out noise and ambiguities in real-world data, especially in the case of gradient-based learning in neural networks \cite{Courbariaux_etal14_lowprec}. 

With this premise, in this chapter, we explore promising learning and inference algorithms compatible with neuromorphic hardware, with a special focus on spiking neural networks. We highlight their hardware requirements, discuss their possible implementations with memristors and finally, identify a class of computations they are particularly suitable for.
In so doing, we will view spiking neural networks as types of recurrent neural networks and explain which hardware non-idealities are detrimental, and which can be exploited for improving computations.

\section{Models of Spiking Neural Networks and Synaptic Plasticity}\label{sec:lif}
We start with a description of neuron models commonly used in neuromorphic hardware and then describe the tools used to analyze and develop algorithms on them. 
The most common model is the \ac{LIF} \cite{Indiveri_etal11_neursili}. While several variations of \ac{LIF} neuron models of it exist, including mixed-signal current-mode and digital implementations, the base model can be formally described in differential form as:
\begin{align}
    \label{eq:ulif_basic}
    \tau_\mathrm{mem} \frac{\mathrm{d}U_i}{\mathrm{d}t} &= -(U_i-U_\mathrm{rest}) + RI_i - S_i(t)(U_i-U_\mathrm{rest}),\\
    \tau_\mathrm{syn} \frac{\mathrm{d}I_i}{\mathrm{d}t} &= -I_i(t) + \tau_\mathrm{syn}\sum_j W_{ij} S_j(t), 
    \label{eq:lif_basic}
\end{align}
where $U_i(t)$ is the membrane potential of neuron $i$, $U_\mathrm{rest}$ is the resting potential, $\tau_\mathrm{mem}$ and $\tau_{syn}$ are the membrane time constants, $R$ is the input resistance, and $I_i(t)$ is the synaptic current \cite{Gerstner_etal14_neurdyna}.
\refeq{eq:ulif_basic} shows that $U_i$ acts as a leaky integrator of the input current $I_i$, which itself is a leaky, weighted integration of the spike train $S_j$. Spike trains are described as a sum of delta Dirac functions $S_j(t) = \sum_k \delta(t-t_j^k)$ $t_j^k$ where $t_j^k$ are spike times of neuron $j$.
In other words, for each incoming spike, the synaptic current undergoes a jump of height ${W_{ij}}$ and otherwise decays exponentially with a time constant $\tau_{\mathrm{syn}}$. Note that the expression $\sum_j W_{ij} S_j(t) $ has the structure of a vector-matrix multiplication, and forms the basis of RRAM implementations of \acp{SNN} discussed later in this chapter. Because $S_j$ represents spikes ($S_j=0$ or $1$), these operations consist in additions and scaling of the result by $\tau_{syn}$.

\begin{figure}
\centering
\vspace{-0.15in}
\subfloat[]{\includegraphics[width=0.5\linewidth]{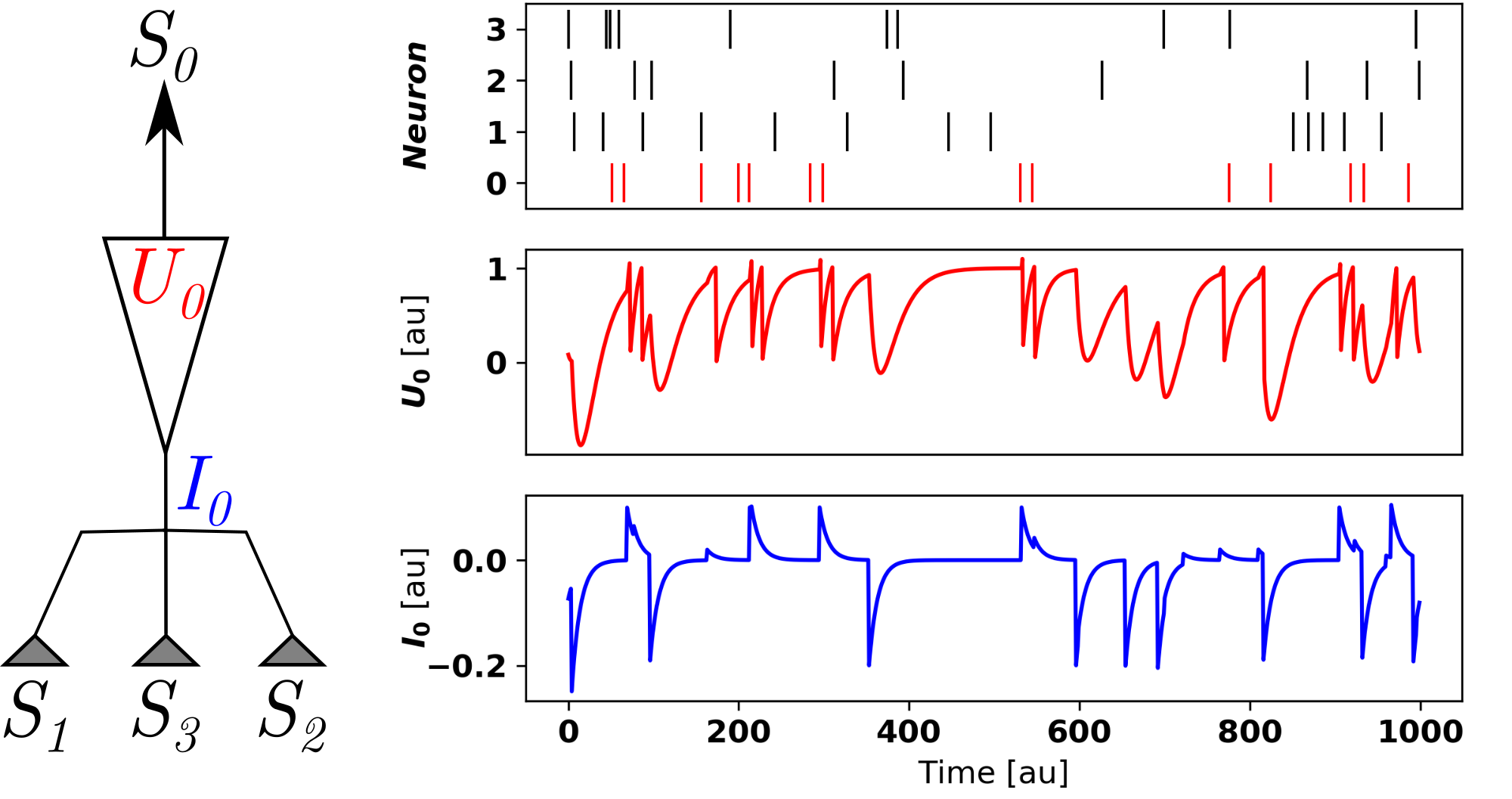}%
\label{fig:lif_neuron}}
\hfil
\subfloat[]{\includegraphics[width=0.45\linewidth]{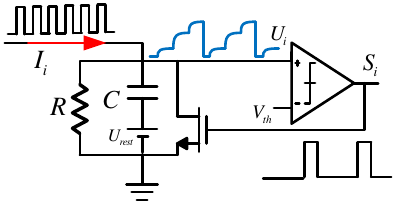}%
\label{fig:lif_circuit}}
  \caption{a)Example \acl{LIF} Neuron dynamics for $U_{rest}=0$ and $V_{th}=1$ and b) basic neuron circuit implementing \ac{LIF} dynamics. }
\label{SP}
\vspace{-0.05in}
\end{figure}

Neurons emit spikes when the membrane voltage reaches the firing threshold, $V_{th}$.
After each spike, the membrane voltage $U_i$ is reset to the resting potential $U_\mathrm{rest}$ (\reffig{fig:lif_neuron}).
In Eq.~\eqref{eq:ulif_basic}, the reset is carried out by the term $S_i(t)(U_i-U_\mathrm{rest})$ when the membrane potential reaches $V_{th}$. This reset acts as a refractory mechanism since the neuron must reach the firing threshold from $U_\mathrm{rest}$ to fire again. 

A simple circuit implementation of the \ac{LIF} is based on resistor-capacitor circuits with a switch to reset the voltage across the capacitor as shown in Fig. \ref{fig:lif_circuit}. The mathematical model of this circuit is equivalent to \refeq{eq:ulif_basic} where $\tau_\mathrm{mem}=RC$. Several variations of this circuit have been demonstrated \cite{Indiveri_etal11_neursili}.
Synaptic circuits which reproduce the dynamics of $I$ can be implemented in analog VLSI using a Differential-Pair Integrator (DPI) or other similar circuits \cite{Bartolozzi_Indiveri07_synadyna}.

To make the mathematical analysis and digital implementations more intuitive, it is useful to write \ac{LIF} in the discrete form:
\begin{align*}
    U_i[n+1] & = \beta U_i[n] + U_{rest} + I_i[n] - S_i[n] (U_i[n] - U_{rest})\\
    I_i[n+1] & = \alpha I_i[n] + \sum_j W_{ij} S_j[n]\\
    S_i[n] &= \Theta(U_i[n])=\left\{ \begin{array}{cc}
 1 & \;U_i[n]\geq V_{th} \\
 0 & \;Otherwise
 \end{array}\right. 
\end{align*}
where $\alpha=\exp(-\frac{\Delta t}{\tau_{\mathrm{mem}}})$, $\beta=\exp(-\frac{\Delta t}{\tau_{\mathrm{syn}}})$ are constants that capture the decay dynamics of states $U$ and $I$ during a $\Delta t$ timestep, $\Theta$ is the step function and $V_{th}$ is the firing threshold.

Several digital implementations of spiking neural networks use these update equations \cite{Detorakis_etal18_neursyna,Davies_etal18_loihneur}.
These equations are can be described using an artificial \ac{RNN} \cite{Neftci_etal19_surrgrad} formalism. 
The recurrent aspect arises from the statefulness of the neurons (when $\alpha>0$ and $\beta>0$). 
In addition, if there exist connections from the neuron to itself, then these connections can be viewed on the same footing as recurrent connections in artificial \acp{RNN}. 
As we shall later see, the artificial \ac{RNN} view of spiking neural networks enables the derivation of biologically credible learning rules with the capabilities of machine learning algorithms.

This description of the spiking neural network generalizes to a non-recurrent artificial neural network where activations are binary. In fact, replacing $\alpha$ and $\beta$ with 0 and ignoring the reset, the equations above become:
\begin{equation}
  \begin{split}\label{eq:perceptron}
    U_i[n+1] & = \sum_j W_{ij} S_j[n],\\
    S_i[n] &= \Theta(U_i[n]).
  \end{split}
\end{equation}
The dynamics above are those of the standard artificial neural network (without any multiplications, as described above) followed by a spiking nonlinearity, \emph{i.e.} they are Perceptrons.
Neural networks with binary neural activations and/or weights were proposed as efficient implementations of conventional neural networks \cite{Courbariaux_etal16_binaneur,Rastegari_etal16_xnorimag}. Such devices are promising for energy-efficient implementations of deep learning processors in full-digital technologies \cite{Andri_etal16_yodaultr,Umuroglu_etal17_finnfram} as well as with in-memory processing with emerging memory devices \cite{Sun_etal18_fullpara}.

\section{ Memristive Realization and Non-idealities}

Neuromorphic hardware implementations require a device or a circuit that mimics the synapse behavior. A minimum requirement for neural network applications is a memory to store the synaptic weight. Memristive devices can be used to realize these such synapses. A device is referred to as a memristive device if it exhibits pinched hysteresis behavior in the current-voltage plane which indicates a memory behavior in its resistance. %
Many physical devices exhibit memristive behaviors such as Phase Change Memory (PCM), ferroelectric RAM (FeRAM), spin-transfer torque magnetic RAM (STT-MRAM), and resistive RAM (RRAM or ReRAM). RRAMs have a promising potential for neuromorphic applications due to high area density, stackability, and low write energy, compared to the other emerging devices \cite{wilson2013international}. Thus, in this section, we focus on RRAM, without loss of generality, to discuss the physical limitations and problems facing deploying such technology for neuromorphic hardware.

\begin{figure}[!h]
\centering
\includegraphics[width=0.8\linewidth]{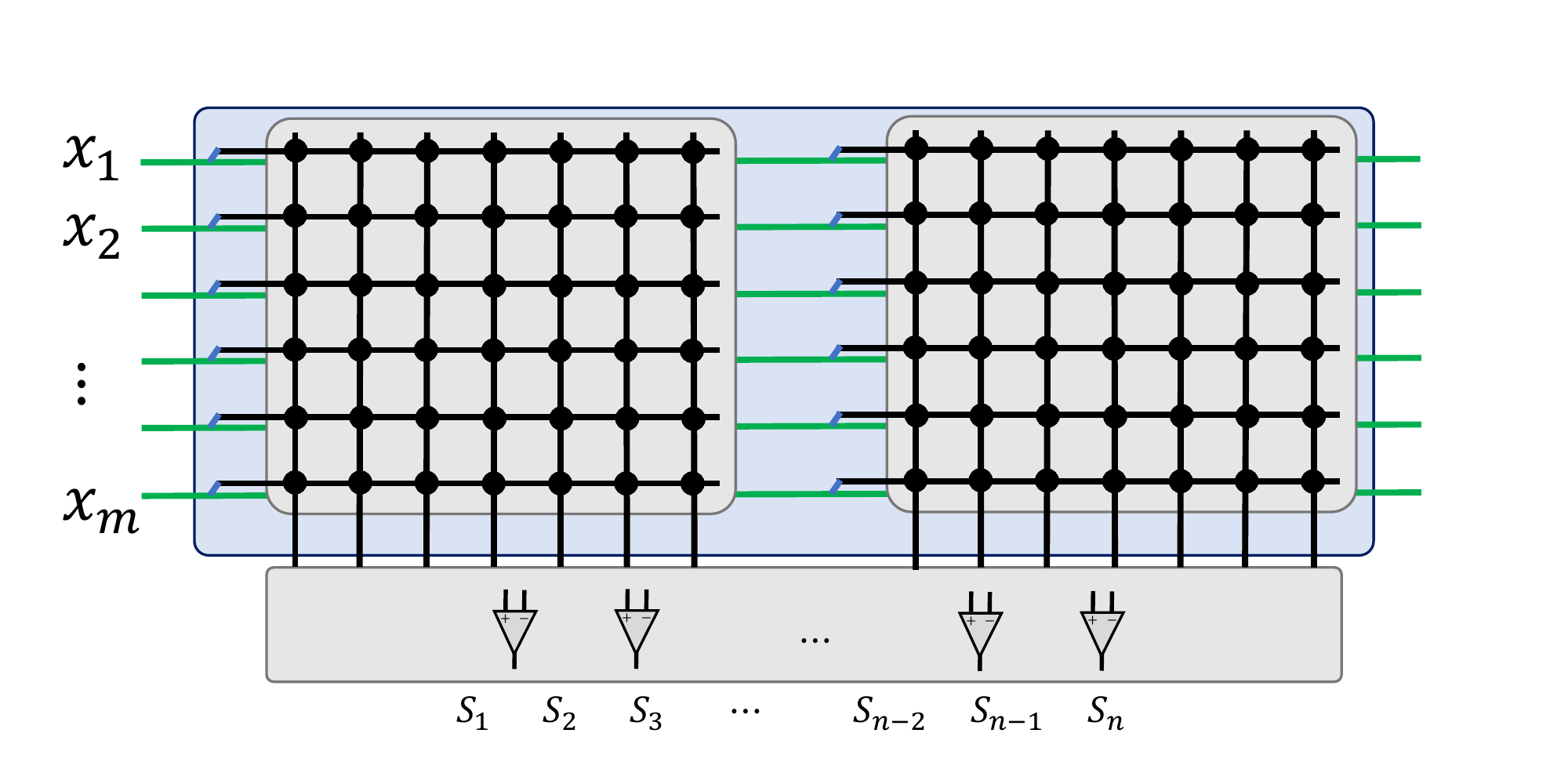}%

\caption{\label{fig:crossbar_array} Crossbar array realization of one layer neural network.}
\end{figure}

A common building block in both spiking and non-spiking neural network is the weighted summation of inputs (see \refeq{eq:perceptron}). This can be performed in a single step using a crossbar structure, unlike the conventional computing methods which typically require $N\times M$ steps or clock cycles. 
\reffig{fig:crossbar_array} shows a single-layer crossbar based resistive neural network with $M$ inputs and $N$ outputs representing $N$ perceptrons with $M$ inputs each and the weights are stored in the memristors. 
The inputs to the perceptrons (presynaptic signals) are encoded in the input voltages, and the output of each perceptron is the sum of the currents passing through each memristor. 
This way, all currents within the same column can be linearly summed to obtain the postsynaptic currents, i.e. the equivalent of \refeq{eq:lif_basic}.
The total postsynaptic currents need sensing and shaping circuits to convert them into voltages and passed to subsequent neurons. In a nonspiking neural network, the postsynaptic current is summed up through the sensing circuit and passed through another shaping circuit to create the required neural activity such as sigmoid, tanh or rectified linear functions.  
With spiking neurons, the output of the current sensing circuit is instead passed through a \ac{LIF} circuit.

In neural networks, both positive (excitatory) and negative (inhibitory) connections are required. 
However, the RRAM conductance is positive by definition which only supports excitatory or inhibitory connections. 
Two weight realization techniques are possible to create both excitatory and inhibitory connections; 

1) using two RRAMs per weight \cite{prezioso2015training,li2018efficient} or

2) using one RRAM as weight in addition to one reference RRAM having a conductance set to $G_r=(G_{max}+G_{min})/2$ \cite{chang2017mitigating,fouda2018overcoming}. 

The first realization has double the dynamic range $w\in [-\Delta G, \Delta G]$, where $\Delta G=G_{max}-G_{min}$, making it more immune to variability at a cost of double the area, double power consumption during reading and additional programming operations. 
The second technique has one RRAM device, meaning that $w\in [-\Delta G/2, \Delta G/2]$ making it more prone to variability but the overall area is smaller, requires less power, and is easier to program (programming only one RRAM per weight).  
Due to the high variations in the existing devices, the first approach is commonly used with either one big crossbar or two crossbars (one for positive weights and the other one for negative weights as shown in \ref{fig:crossbar_array}). 
The output of the memristive neural network can be written as
\begin{equation}
    S_i=\sum_{j=1}^m G_{ij} V_j
\end{equation}
\noindent where $S_i$ is the output of the $i^{th}$ neuron and $G_{ij}=G_{ij}^{+} - G_{ij}^{-}$, is the synaptic weight, and $V_j$ is the $j^{th}$ input. 
The crossbar array forms the majority of the research in using RRAMs for neuromorphic computation \cite{yu2018neuro}.
In practice, RRAMs and crossbar structures suffer from many problems and do not behave ideally for computational purposes.
These non-idealities can severely undermine the overall performance of applications unless they are taken into consideration while the training operation. 
After defining the mapping of synaptic weights to RRAM conductances, the following sections will overview these non-idealities in the light of neuromorphic computation and learning.
\subsection{Weight Mapping}
\begin{figure}[!h]
\centering
\includegraphics[width=0.35\linewidth,height=0.4\linewidth]{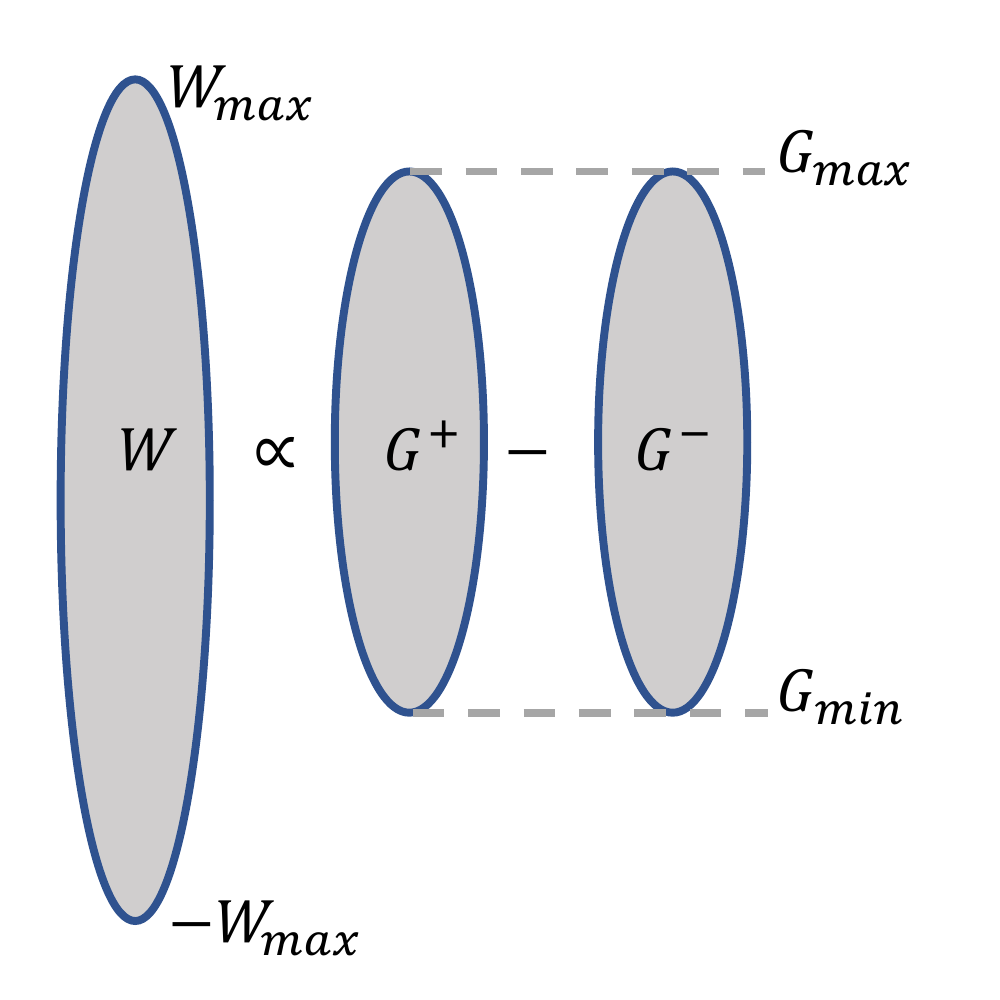}
\caption{\label{fig:mapping}Mapping synaptic weight into conductances.}
\end{figure}
As discussed above, each weight is translated into two conductances which is one-to-two mapping which can be mathematically formulated as 
\begin{equation}
    G=G^+-G^-=\frac{W}{W_{max}}\Delta G,
\end{equation}
\noindent where $W_{max}$ is the maximum value of the weight. If it is required to realize $W_{max}$, $G^+$ and $G^-$ are set to $G_{max}$ and $G_{min}$, respectively.   
The difference between the two conductances is constant and proportional to the required weight value and each conductance is constrained to be between $G_{min}$ and $G_{max}$ as shown in \reffig{fig:mapping}. Thus, there are many possible realizations for each weight; for example, the zero weight can be realized with any equal values of $G^+$ and $G^-$. Therefore, a criterion for selecting the weight mapping should be defined. This can be formulated as an optimization problem as follows:
\begin{equation}
\begin{aligned}
  & {\text{minimize}} \quad \quad \mathcal{L}(\textsc{G}^+, \textsc{G}^-) \\
&\text{subject to} \\
&  \quad \quad  \quad \quad  G_{min}\leq \textsc{G}^+ ,\quad \textsc{G}^- \leq G_{max}, \text{and} \\ 
&  \quad \quad  \quad \quad \textsc{G}^+ - \textsc{G}^-=\frac{\textsc{W}}{W_{max}}\Delta G.
\end{aligned}
\end{equation}
\noindent where $\mathcal{L}$ is the objective function (objective function are discussed in further detail later in this chapter). $\textsc{G}^+$ and $\textsc{G}^-$ are the positive and negative conductance matrices.   
Since many conductance configurations are possible to obtain the same effective weight, additional criteria such as power consumption while reading (important for inference) or the write energy (important in online training) can be introduced.
These constraints can be taken into consideration while training the network with a regularization term in the loss function (see \refsec{sec:learning}). 
We note that the mapping is more important for offline training where the optimization is completed on software and the final weight values are transferred to the hardware.
\subsection{RRAM endurance and retention}
An attractive purpose of RRAMs is to accelerate inference and training.
However, endurance is a critical obstacle to RRAM deployment in neuromorphic hardware. 
In online learning, the devices are frequently updated, and especially so during gradient-based learning such as in artificial neural networks.
However, each device has a limited number of potentiation and depression cycles \cite{zhao2018characterizing,chen2011physical}. 
Endurance depends on the device's switching and electrode materials. For example, $\mathrm{HfO_x}$ devices can achieve endurance up to $10^{10}$ cycles, but $\mathrm{Al_2O_3}$ devices achieve endurance up to $10^4$ \cite{nail2016understanding}.  
With limited endurance, it is necessary to complete the training before the devices degrade.
The endurance requirement for learning is application-dependent. 
In standard deep learning, weight updates are usually performed every batch. 
Classification benchmarks such as MNIST handwritten digit recognition require writing around $10^4$ cycles.
However, even gradient-based training of neural networks can easily scale to $10^8$ cycles. 
 
Furthermore, because neuromorphic hardware can be multi-purpose (i.e. the same device and be used to perform many different tasks), where a complete training of the network is performed for every task. Consequently, the device endurance should be high enough to cover its lifetime use. 
There are some solutions to mitigate the endurance problem in machine learning scenarios: 
\begin{itemize}
  \item Full offline training: training is completed off the device and the final weights are transferred to the RRAM-based hardware. This requires an accurate model of the devices, the crossbar array and the sensing circuitry in the training procedure, and to verify the response of each part of the network to make sure that the response matches the simulated one \cite{jain2018rx}.  
  \item Semi-online training: A complete training cycle is performed offline, then the new weights are transferred to the devices. Then, online retraining is carried out to reduce performance loss due to the existing impairments. Due to the smaller number of writing cycles, this solution would relax the endurance requirements. In \cite{fouda2018overcoming}, it was noticed that the network was able to recover the original accuracy after $10\sim20\%$ of the training epochs. 
\end{itemize}
Once the online or the offline training is performed, the network can operate in the inference mode where only reading cycles are performed. In this case, the retention of the stored values becomes an important issue. As with endurance, RRAM retention is also dependent on the device materials and temperature. 
For example, the $\mathrm{HfO_x}$ devices have around $10^4$ seconds (2.78 hours) retention \cite{azzaz2016endurance}. 
Although this might be sufficient for certain single-use scenarios, such as biomedical applications, it is inadequate for IoT and autonomous control applications. 
There, retention values need to be more than $10^6$ seconds across different temperature values (since retention degrades with increasing the temperature).  

Full online learning requires high endurance and moderate retention, but, semi-online requires moderate endurance and retention.
Thus, while both endurance and retention are important for machine inference and learning tasks, the learning approach may require one more over the other.

\subsection{Sneak Path Effect}
\begin{figure}[!t]
\centering
\vspace{-0.15in}
\subfloat[]{\includegraphics[width=0.48\linewidth,,height=0.4\linewidth]{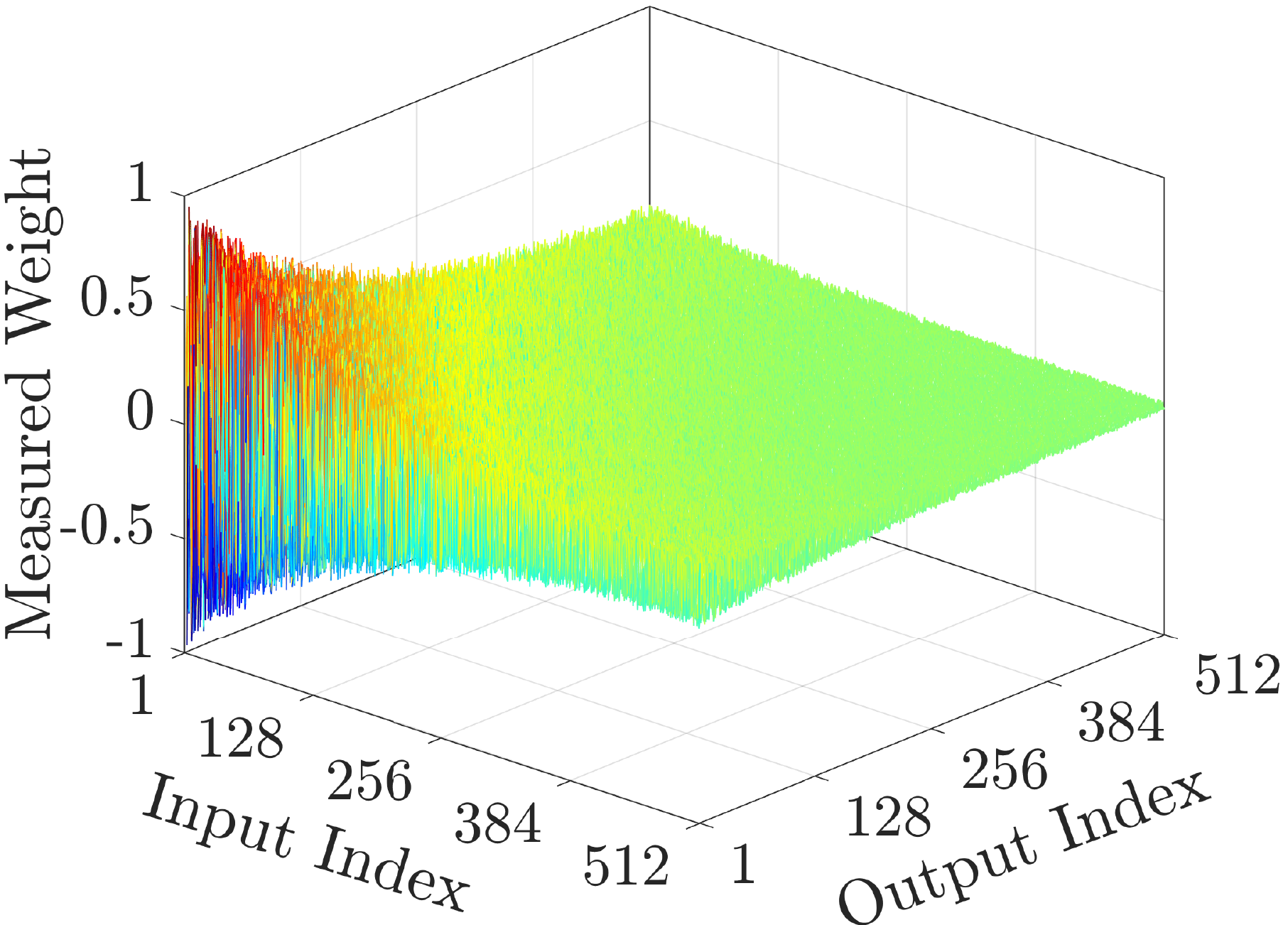}%
\label{SP_3D}}
\hfil
\subfloat[]{\includegraphics[width=0.48\linewidth,,height=0.4\linewidth]{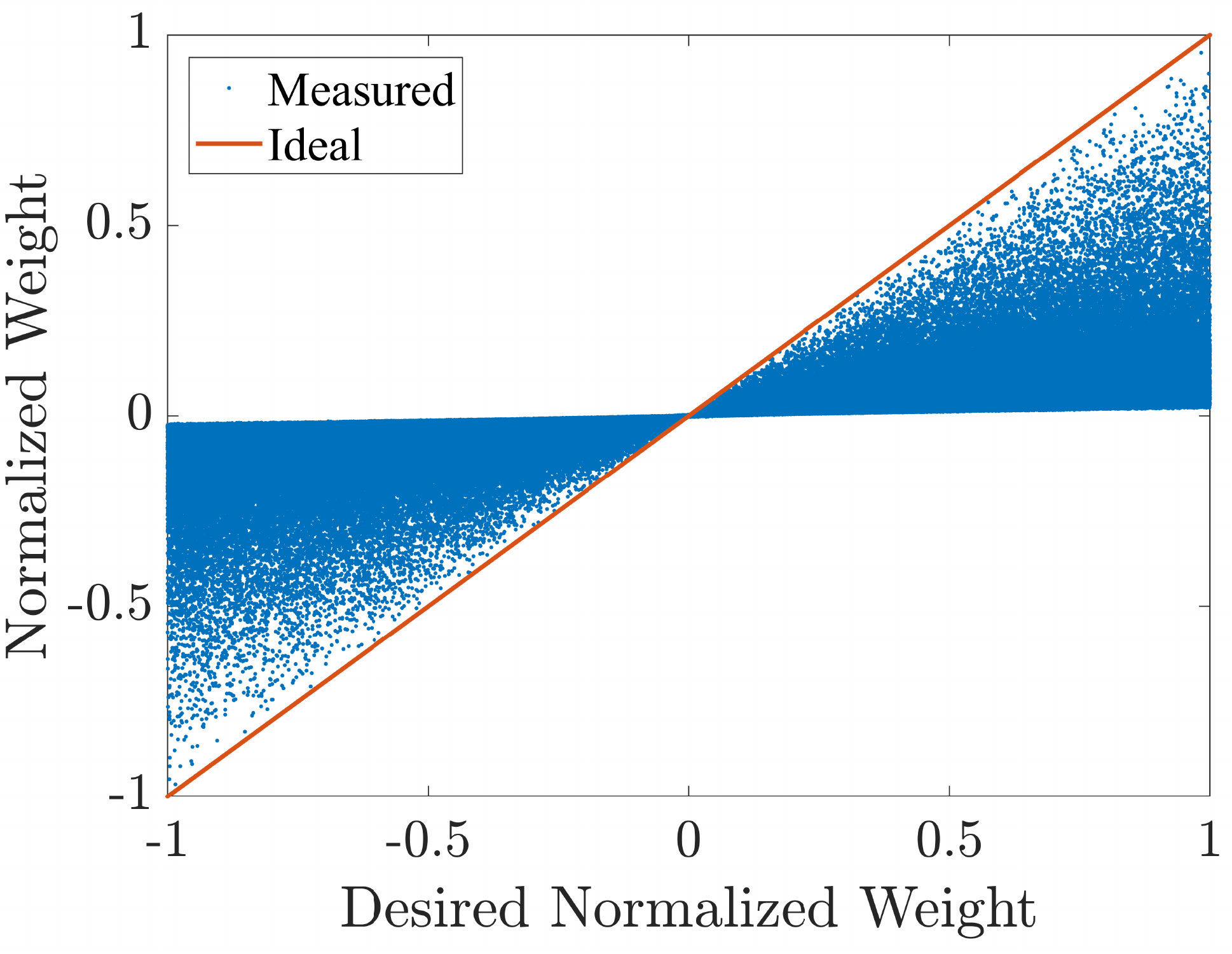}%
\label{SP_}}
\hfil
\subfloat[]{\includegraphics[width=0.48\linewidth,,height=0.4\linewidth]{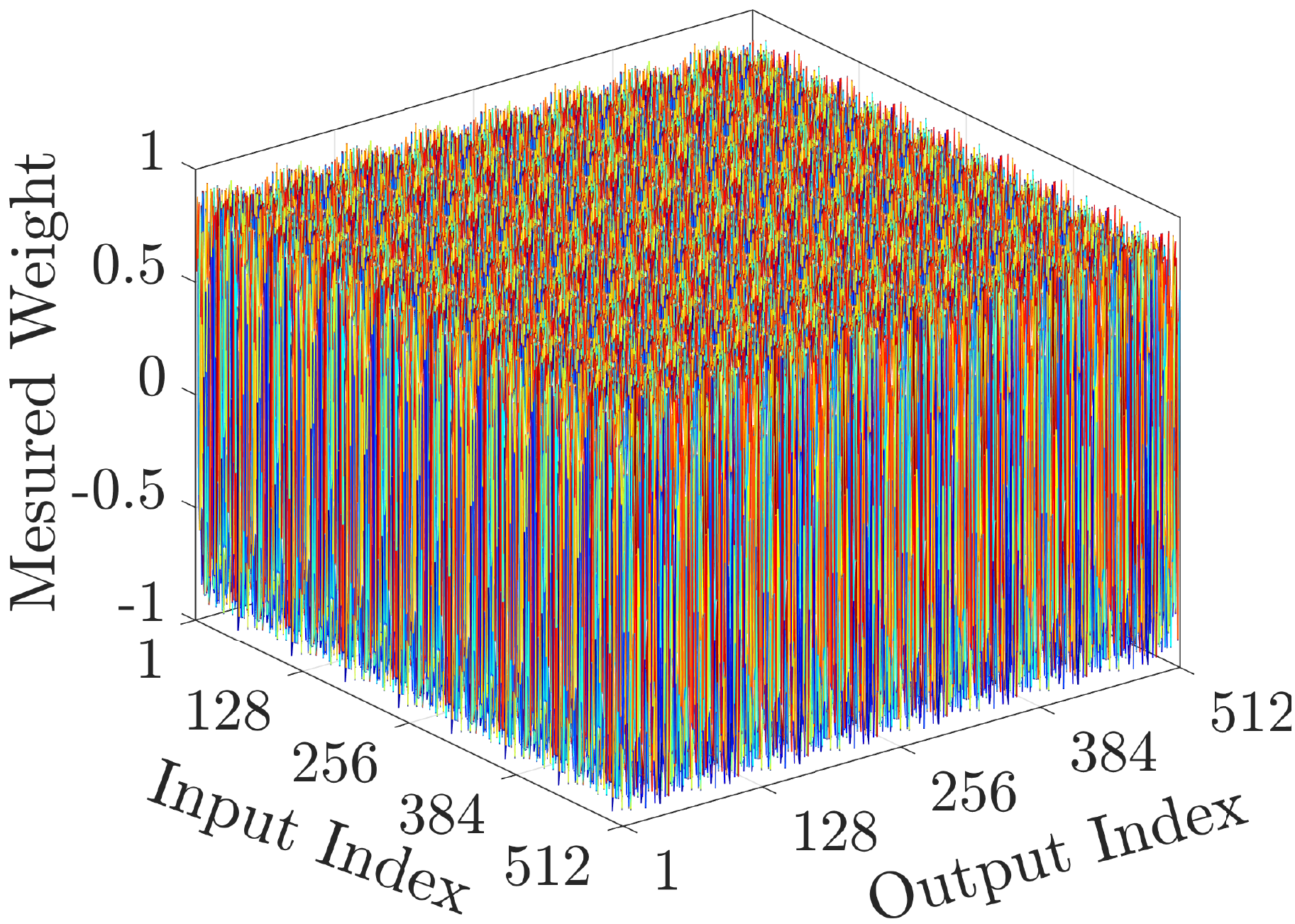}%
\label{SP_3D_64}}
\hfil
\subfloat[]{\includegraphics[width=0.48\linewidth,,height=0.4\linewidth]{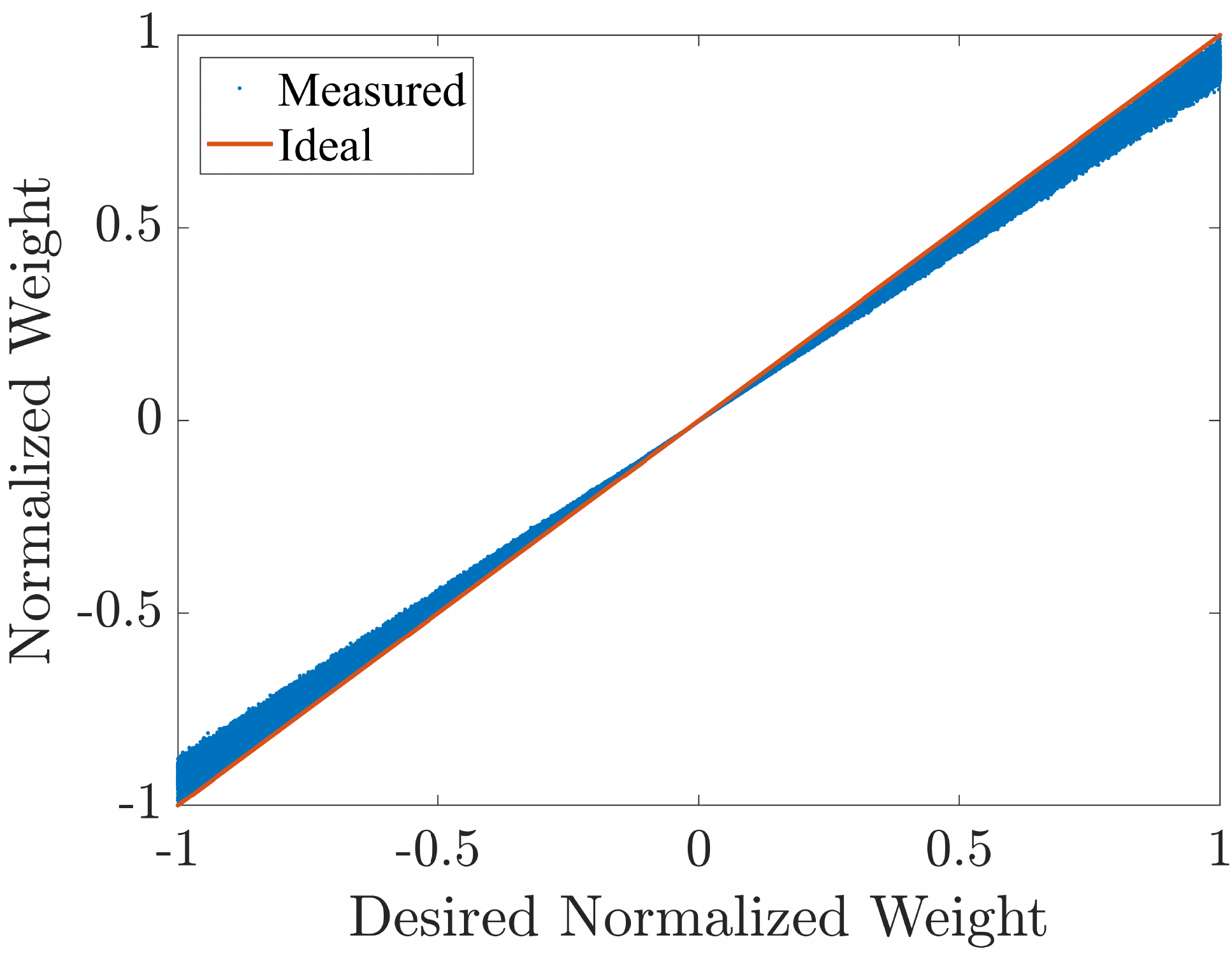}%
\label{SP_64}}
\vspace{-0.05in}
\caption{Effect of the wire resistance on the measured weights for $512\times 512$ crossbar array at with $0.1\Omega$ wire resistance. 3D plots of random weights distributed across the array, (a) without partitioning and (c) with partitioning into $64\times 64$ crossbar arrays; and the measured weights with the sneak path problem versus the desired values for (b) the entire array without partitioning and (d) with partitioning.}
\label{SP}
\vspace{-0.05in}
\end{figure}

The sneak path problem, also is referred to as IR drop, arises from the existence of the wire resistance which is inevitable in nanostructure crossbar arrays. The wire resistance creates many paths to the signal from each the input port to the output port. These multiple paths create undesired currents which perturb the reading of the weight. It is expected that the wire residence would reach around $92\Omega$ for $5nm$ feature size \cite{fouda2018modeling}, which is the expected feature size for crossbar technology according to International Technology Roadmap of Semiconductors (ITRS) \cite{wilson2013international}. \reffig{SP_3D} and \reffig{SP_} show an example of the sneak path problem in $512 \times 512$ with random weights. A linear switching device having a $10^6\Omega$ high resistance state and $10^3\Omega$ low resistance state is used while the wire resistance is $0.1\Omega$. Ideally, the measured weights should be similar to the measured weights, as shown in Fig. \ref{SP_3D_64}. Despite the small value of the wire resistance, it has a very high effect on the weights stored in the crossbar arrays (\reffig{SP_3D}). The weights are exponentially decaying across the diagonal of the array where the cell (1,1) has the least sneak path effect and the cell (N,M) has the worst sneak path effect. 

Some devices have a voltage-dependent conductance where the conductance is an exponential or quadratic function of the applied voltage \cite{fouda2018modeling}. 
This conductance nonlinearity can help reduce the sneak path problem in resistive memories on crossbar or crosspoint structures \cite{fouda2019resistive} due to single cell reading. 
But, in neuromorphic applications, this adds an exponential behavior to vector-matrix multiplication (VMM) which becomes
\begin{equation}
    S_i=\sum_{j=1}^m G_{ij} \sinh(a V_j). 
\end{equation}
\noindent This exponential nonlinearity makes the VMM operation inaccurate which deteriorates the training performance \cite{kim2018deep}. Some algorithms were developed to take the effect of the device's voltage-dependency into consideration while training non-spiking neural networks such as \cite{kim2018deep}. The same algorithm idea can be extended to spiking neural networks.

\begin{figure}[!h]
\centering
\includegraphics[width=\linewidth]{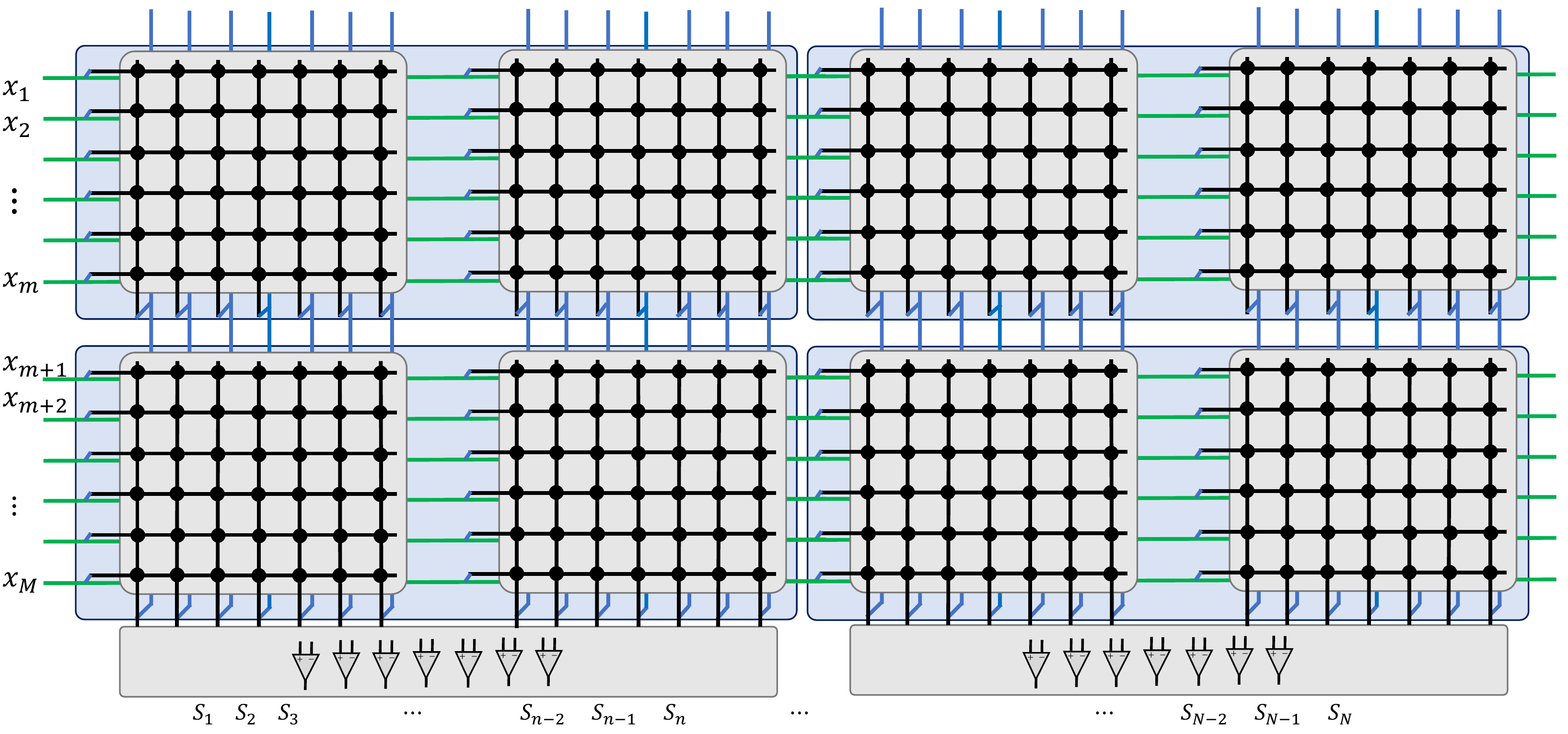}%
\caption{Realization of the partitioned matrices. 
\label{fig:partioning}
  }
\end{figure}

\paragraph{Partitioning of Large Layer Matrices}
The sneak path problem prohibits the implementation of large matrices using a single large crossbar array. One possible solution is to partition the large layer matrices into small matrices that can be implemented using realizable crossbar arrays. Figure \ref{fig:partioning} shows the partitioned crossbar arrays and the interconnect fabric between them to realize the complete VMMs where the large crossbar array, having $N\times M$ RRAMs, is partitioned into $n\times m$ crossbar arrays.
In order to have the same structure of a large crossbar array, vertical and horizontal interconnects are placed under the crossbar arrays. This horizontal interconnect is used to connect the inputs between the crossbar arrays within the same array rows. The vertical interconnect is used to connect the outputs of the vertical crossbar arrays. The vertical interconnects are grounded through the sensing circuit to absorb the currents within the same vertical wire. The sensed currents are connected then to the neuronal activity. It is worth highlighting that each crossbar array may require input drivers (buffers) to reduce the loading effect of the vertical interconnect and crossbar arrays. These buffers are not shown in \reffig{fig:partioning} for clarity.  Moreover, they can be placed under the crossbar arrays to save the wafer area where the crossbar arrays are usually fabricated between higher metal layers.   %
\reffig{SP_3D_64} shows the measured random synaptic weights with the same aforementioned parameters after partitioning the $512\times 512$ crossbar arrays into 64 of $64\times 64$ crossbar arrays. The weight variations due to their locations in the crossbar array became much smaller as shown in \reffig{SP_64} and can be considered with the device variation.

Although partitioning the array mitigates the sneak path problem, it might cause routing problems where the non-idealities (e.g parasitics) of the routing fabric will affect the performance. Thus, routing's non-idealities must be simulated in the case of full-offline learning. Also, additional algorithmic work is needed to overcome the residual sneak path problem after partitioning (especially with the aforementioned high wire resistance expected to be $10\Omega$ per cell), such as the masking technique proposed in binary neural networks \cite{fouda2018overcoming}. In the masking technique, the exponentially decaying profile is used to capture the effect of the sneak path problem during learning by multiplying element-wise the trained weights. 

\subsection{Delay}
Signal delay determines the speed at which computations can be run on hardware. While delays are not an issue for neuromorphic hardware designed to run with real-work time constants \cite{Benjamin_etal14_neurmixe,Qiao_etal15_recoon-l}, other models are accelerated \cite{Schemmel_etal08_wafeinte}. Due to the parallel VMM operation, the memristive hardware would be dedicated to an accelerated regime. For this, it is necessary to reduce delay is caused by the device and structure parasitics and circuits.

In \cite{fouda2018modeling}, a complete mathematical model for the crossbar delay is discussed. The model showed that the delay is a function of the weights stored in the crossbar arrays. The higher the device resistance, the more delay to the signal. For $1M\Omega$ switching resistance, the maximum delay of the crossbar arrays is expected to be in the range of nanoseconds. Also, there is another delay resulting from the sensing circuit which is expected to be around $10$ns.

The partitioning and the drivers add extra delay factors can be caused by the wire resistance of the interconnect fabric and the input capacitance of the drivers. 
Delay can be calculated using the Elmore delay model \cite{fouda2018modeling}. 
The wire resistance of the interconnect per array is $nR_w$ where $n$ is the number of columns per array and $R_w$ is the wire resistance per cell. 
The Elmore delay of such an interconnect wire is $0.67nR_wC_d$, where $C_d$ is the input capacitance of the buffer. Thus the total input delay is $0.67(N-n)R_wC_d+(N/n)\tau_d$, where $N/n$ is the number of horizontal crossbar arrays and $\tau_d$ is the driver delay. The delay resulting from the partitioning and drivers is expected to be in the range of nanoseconds. Thus, the total delay of the entire layer would be in the range of $20-100\,$ns. %
It is worth mentioning that the effect of the capacitive parasitics of the crossbar array is often ignored because the feature size of the fabricated devices is in the range of sub micrometers, i.e. $F=200nm$, \cite{Prezioso_etal15_traioper}. However, for nano-scale structures, i.e. $F=10nm$, the capacitive parasitics may cause leakage paths at high frequency, where the impedance between the interconnects would be comparable to or less than the switching device's impedance, which would affect the performance. Thus, a more detailed analysis of the capacitive parasitics of the crossbar array must be considered on a case-to-case basis.

\subsection{Asymmetric Nonlinearity Conductance Update Model}
Several RRAM devices demonstrating promising synaptic behaviors are characterized by nonlinear and asymmetric update dynamics, which is a major obstacle for large-scale deployment in neural networks \cite{yu2018neuro}, especially for learning tasks. Applying the vanilla backpropagation algorithms without taking into consideration the device non-idealities does not guarantee the convergence of the network. Thus, a closed-form model for the device nonlinearity must be derived based on the device dynamics and added to the neural network training algorithm to guarantee the convergence to the optimal point (minimal error). 

Most of the potentiation and depression behaviors have exponential dynamics versus the programming time or the number of pulses. 
In practice, the depression curve has a higher slope compared to the potentiation curve, which causes asymmetric programming.
The asymmetric nonlinearity of the RRAM's conductance update can be fitted to the following model 
\begin{equation}
G(t)= \left\{ \begin{matrix}
G_{max}-\beta_P e^{-\alpha_1 \phi(t)} & v(t)>0 \\
G_{min}+\beta_D e^{-\alpha_1 \phi(t)} & v(t)<0\\
\end{matrix}
\right.
\label{ANM}
\end{equation}
\noindent where $G_{max}$ and $G_{min}$ are the maximum and minimum conductances respectively, $\alpha_1, \alpha_2, \beta_P$ and $\beta_D$ are fitting coefficients. $\beta_P$ and $\beta_D$ are related to the difference between $G_{max}$ and $G_{min}$ and $\phi(t)$ is the time integral of the applied voltage. 

Updating the RRAM conductance is commonly performed through positive/negative programming pulses for potentiation/depression with pulse width $T$ and constant programming voltage $V_p$. 
As a result, the discrete values of the flux are $\phi(t=nT)=V_p n T$ where $n$ is the number of applied pulses. This technique provides precise and accurate weight updates. For $t=n\Delta T$, and substituting back into \refeq{ANM}, the potentiation and depression conductances become:
\begin{eqnarray}
G_{LTP}=G_{max}-\beta_P e^{-\alpha_P n},\,\, \text{and}\\
G_{LTD}=G_{min}+\beta_D e^{-\alpha_D n},
\end{eqnarray}
\noindent respectively, where $n$ is the pulse number, $\alpha_P=|V_p|\alpha_1 T$ and $\alpha_D=|V_D| \alpha_2T$. 
The rate of change in conductance with respect to $n$ becomes
\begin{equation}
\frac{dG}{dn}= \left\{ \begin{matrix}
\beta_P \alpha_P  e^{\alpha_P n} ,\text{ for LTP}\\
-\beta_D \alpha_D  e^{\alpha_D n} ,\text{ for LTD}\\
\end{matrix} .
\right.
\end{equation}
One way to quantify the device potentiation and depression asymmetry and linearity is the asymmetric nonlinearity factors \cite{woo2018resistive}.
The effect of these factors is reflected in the coefficients $\alpha_P, \alpha_D, \beta_P$ and $\beta_D$ which are used for the training. 
The potentiation asymmetric nonlinearity (PANL) factor and depression asymmetric nonlinearity (DANL) are defined as
$PANL={G_{LTP}\left(N/2\right)}/{\Delta G}-0.5$ and $DANL=0.5-{G_{LTD}\left(N/2\right)}/{\Delta G}$, respectively, where $N$ is the total number of pulses to fully potentiate the device. $PANL$ and $DANL$ are between $[0, 0.5]$. The sum of both potentiation and depression asymmetric nonlinearities represents the total asymmetric nonlinearity (ANL) which can be written as follows for the proposed RRAM model:
\begin{equation}
ANL=1-\frac{\beta_P e^{-0.5\alpha_P N}+\beta_D e^{-0.5\alpha_D N}}{\Delta G}. 
\end{equation}

\subsubsection{Asymmetric Non-linearity Behavior Example}
\begin{figure}[!t]
\centering
\vspace{-0.15in}
\subfloat[]{\includegraphics[width=0.48\linewidth,,height=0.4\linewidth]{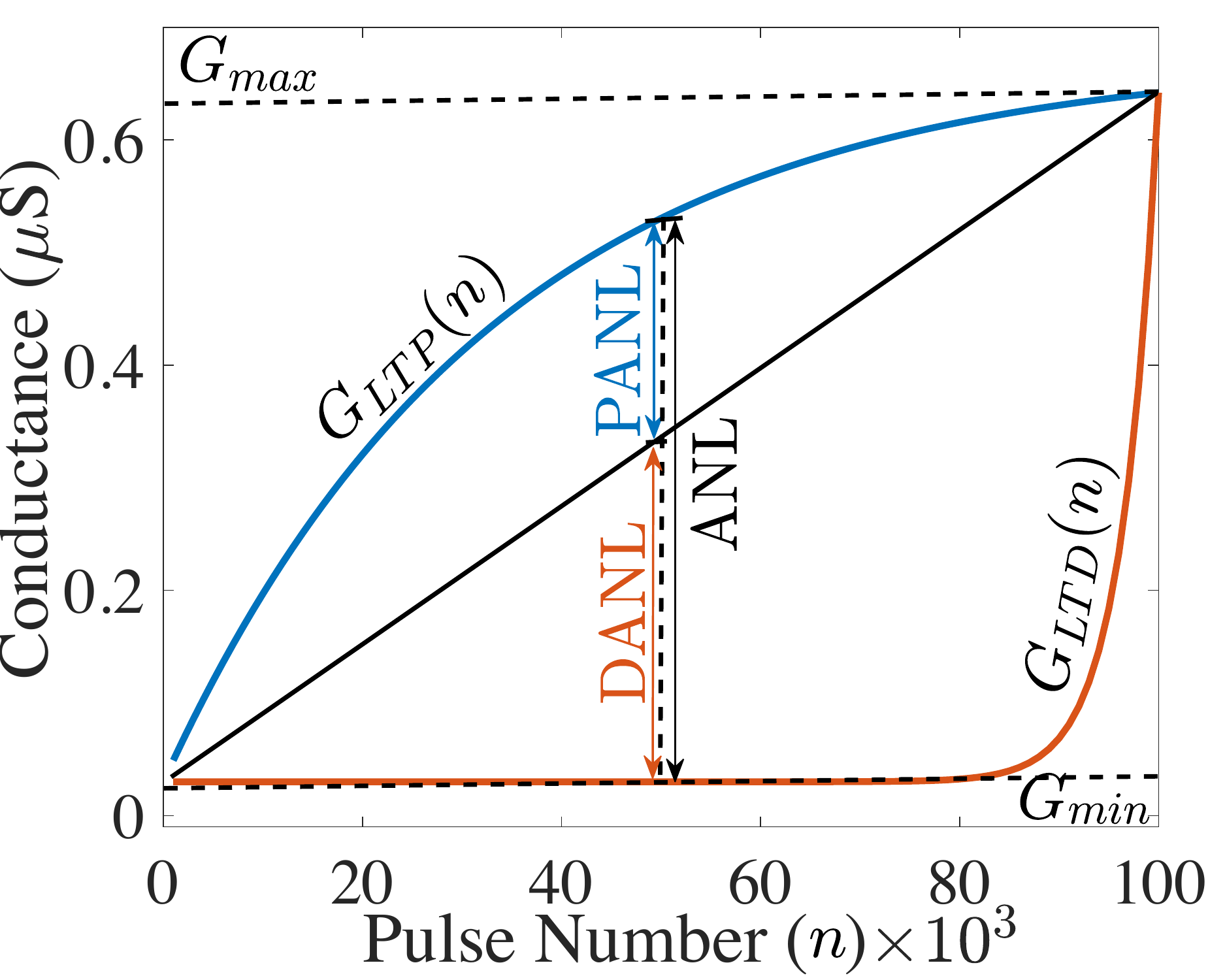}%
\label{ConductanceVsPulses}}
\hfil
\subfloat[]{\includegraphics[width=0.51\linewidth,,height=0.4\linewidth]{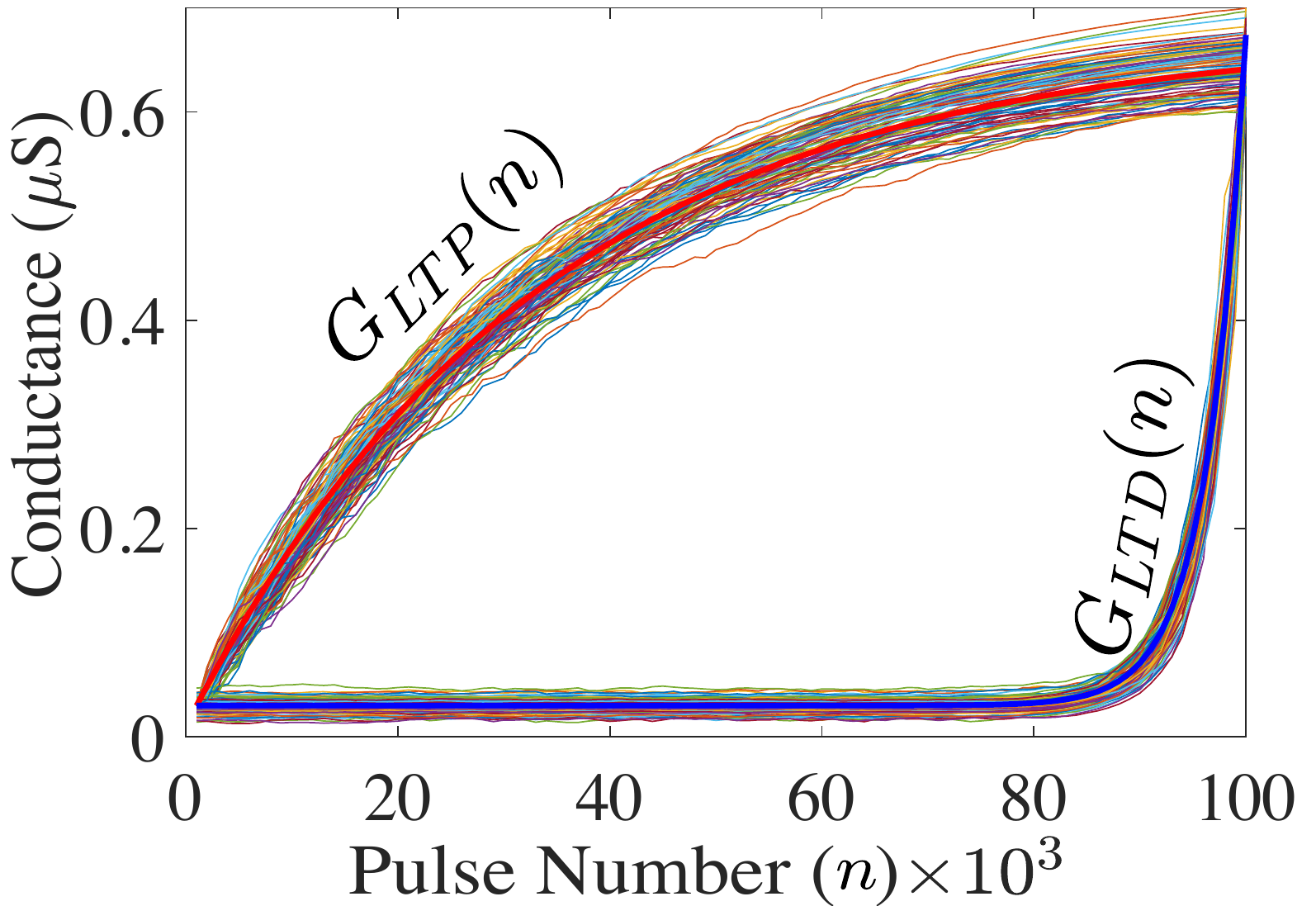}%
\label{ConductanceVsPulses_Variability}}
\vspace{-0.05in}
\caption{Non-idealities of the RRAM:(a) asymmetric nonlinear weight update (b) device Variations. Reproduced from \cite{fouda2018independent}.}
\label{ConductanceVsPulsesFigure}
\vspace{-0.05in}
\end{figure}

An example of a synaptic device is a non-filamentary (oxide switching) ${TiO_x}$ based RRAM with a precision measured to 6 bits \cite{park2016tio}. The ${Mo/TiO_x/TiN}$ device was fabricated based on a redox reaction at ${Mo/Tio_x}$ interface which forms conducting ${MoO_x}$.
This type of interface based switching devices exhibits good switching variability across the entire wafer and guarantees reproducibility \cite{park2016tio}.
The asymmetric nonlinear behavior of this device is shown in Fig. \ref{ConductanceVsPulses}. 

The proposed model was fitted and parameters were extracted for the three programming cases $\{\pm 2V,$ $\pm 2.5V,$ and $\pm 3V\}$. Tables \ref{tab:pot_param} and \ref{tab:dep_param} show the extracted model identification parameters of the device for the three reported voltages with negligible root mean square errors (RMSE). According to the results, the higher the applied voltage, the higher the switching range. Clearly, the model parameters are a function of the applied voltage.
Thus, each parameter can be modeled as a function of the applied voltage which would help to interpolate potentiation and depression curves if non-reported responses are required to be tested. The interpolated models are reported in the tables as functions of the applied voltage.

\begin{figure}[!t]
\centering
\vspace{-0.15in}
\subfloat[]{\includegraphics[width=0.48\linewidth,,height=0.4\linewidth]{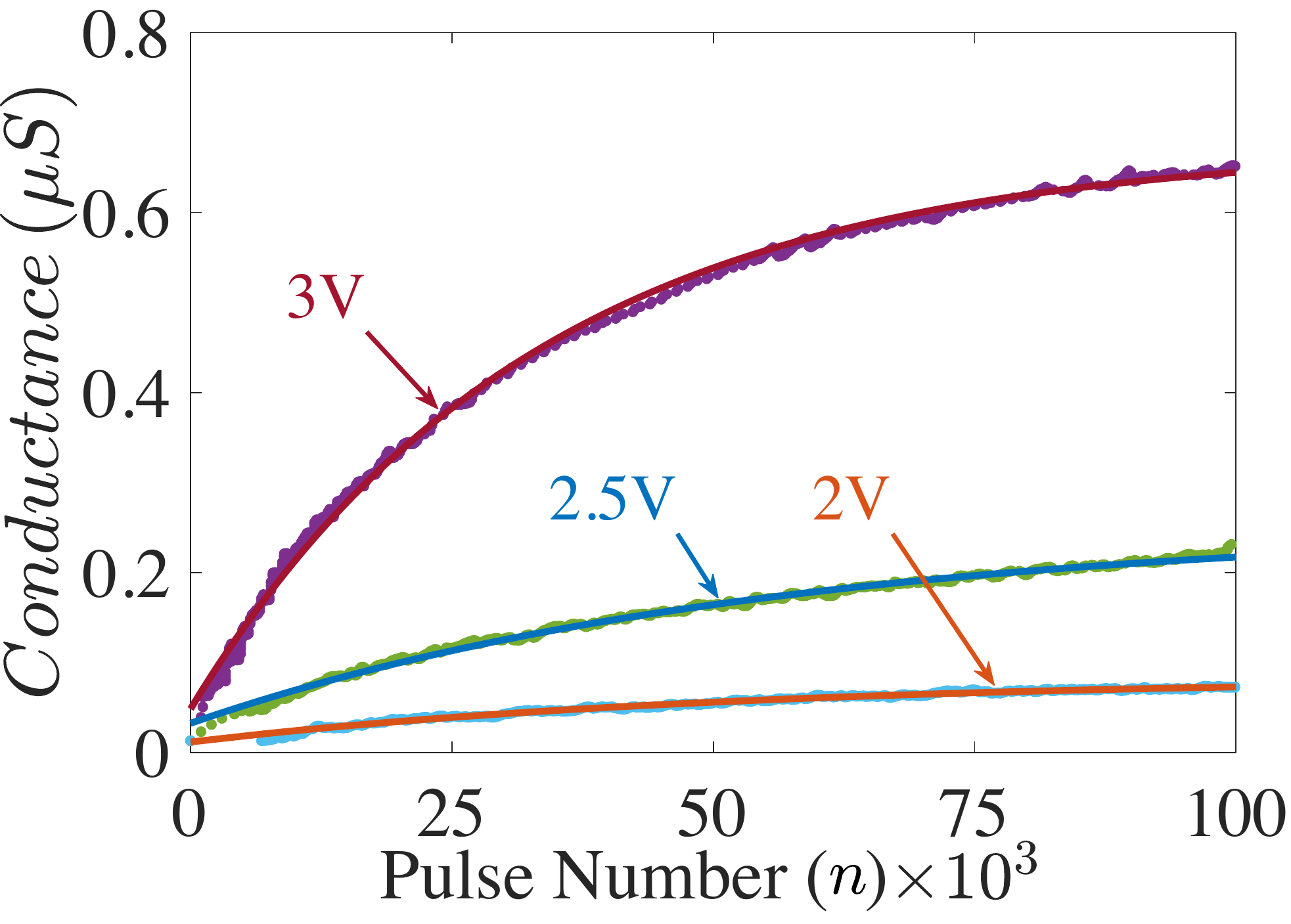}%
\label{FigLTP}}
\hfil
\subfloat[]{\includegraphics[width=0.51\linewidth,,height=0.4\linewidth]{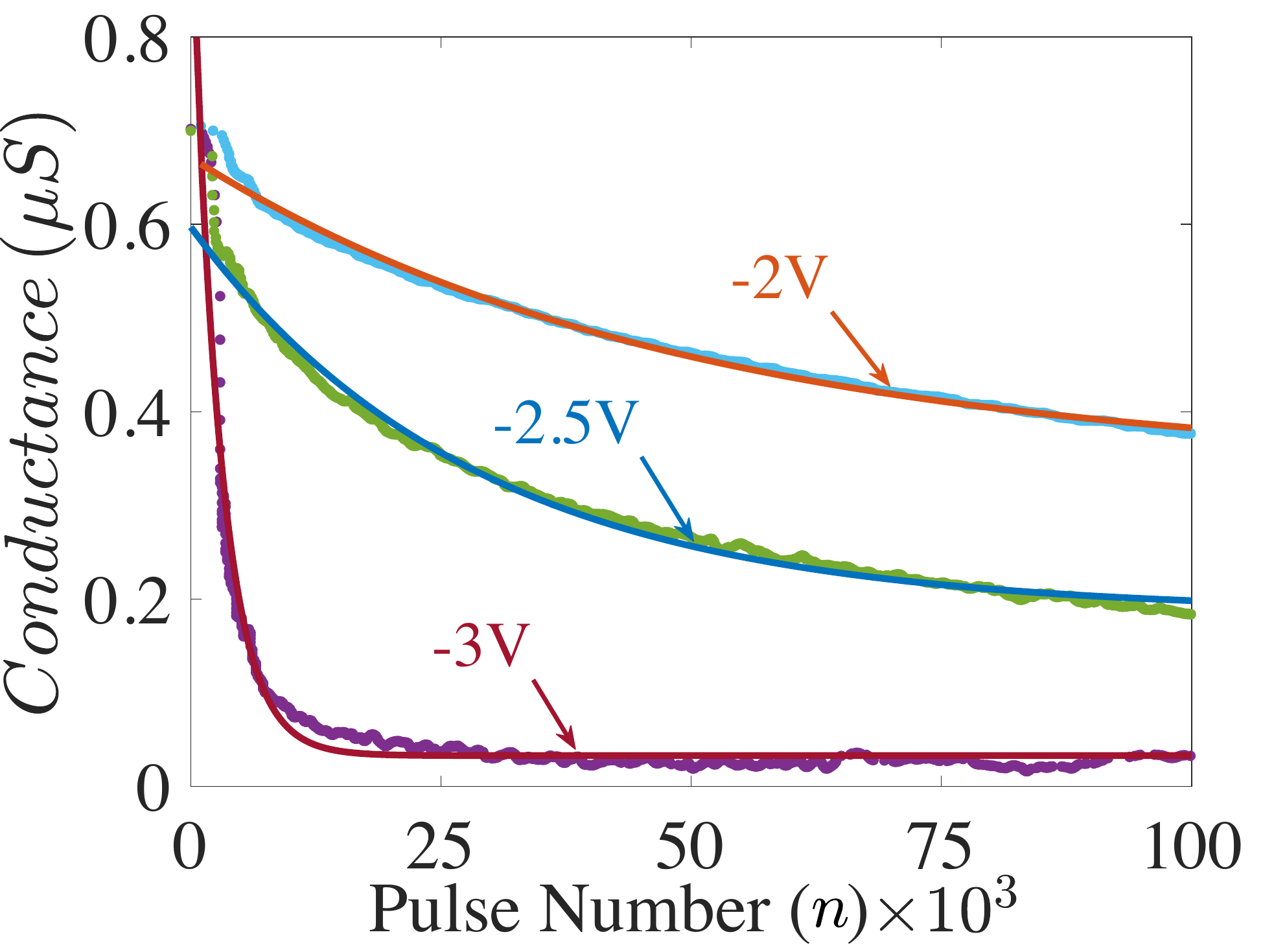}%
\label{FigLTD}}
\vspace{-0.05in}
\caption{RRAM's conductance update (a) long term potentiation  (b) long term depression. Reproduced from \cite{fouda2018independent}.}
\label{FigCond}
\vspace{-0.05in}
\end{figure}

Practically, $V_p=\pm 3V$ cases would be considered since it has the widest switching range. Figure \ref{FigCond} shows the curve fitted model on the top of the reported conductance for both potentiation and depression scenarios. This device has $PANL=0.32$ and $DANL= 0.45$ with $ANL=0.77$. 

\begin{table}[!h]
    \centering
    \begin{tabular}{|c|c|c|c|c|}
    \hline
    $V_p (V)$& $G_{max}$ (nS)& $\alpha_P\times 10^{-3}$ & $\beta_P\times 10^{-9}$& RMSE\\ \hline 
    3& 674&30.58&626.8&9.07  \\  \hline
    2.5& 252.7&18.23&220.22&0.6416 \\   \hline
    2&83.38&19.19&71.7&0.2276   \\ \hline
    $V_p$&$2.968 e^{1.823V_p}-30.4$& 
         $2.019\times 10^{-9} e^{7.51V_p}+18.28$ 
         &$1.522e^{2.014V_p}-13.78$ &
         $-$\\ \hline
    \end{tabular}
    \caption{Extracted Potentiation parameters of the $Mo/TiO_x/TiN$ device extracted from \cite{park2016tio}.}
    \label{tab:pot_param}
\end{table}

\begin{table}[!h]
    \centering
    \begin{tabular}{|c|c|c|c|c|}
        \hline 
         $V_p (V)$& $G_{min}$ (nS)& $\alpha_D \times 10^{-3}$ & $\beta_D\times 10^{-9}$& RMSE  \\  \hline
         -3&32.95&353.4&921.9&23.696  \\   \hline
         -2.5&186.3&35.29&410.9&10.3215 \\   \hline 
         -2&340.5&20.55&330.8&6.12 \\   \hline
         $V_p$&$307.6V_p+955.5$& 
         $8.14\times 10^{-6} e^{-5.48V_p}+20.5$ 
         &$0.009e^{-3.706V_p}+315.9$ &
         $-$\\ \hline
    \end{tabular}
    \caption{Extracted depression parameters of the $Mo/TiO_x/TiN$ device extracted from \cite{park2016tio}.}
    \label{tab:dep_param}
\end{table}
Device variations are an important issue to be taken into consideration during training.
In RRAMs, there are two types of variations: 
(1) The variation during the write operation where a slightly different value is written in the device because of the randomness in the voltage variation and switching materials.
This randomness can be mitigated with write-verify techniques where the written value is read to verify the value and corrected until the desired value is obtained \cite{puglisi2015novel}.
(2) Independent device-to-device due to fabrication mismatch and material inhomogeneity.
These variations can be included in the model by treating each parameter in the model as an independent random variable. Figure \ref{ConductanceVsPulses_Variability} shows the conductance variations of multiple devices during the potentiation and depression cycles with $\pm 3V$ programming pulses. 
The model parameters are sampled from Gaussian sources with $25\%$ tolerance (Variance/mean) for $\alpha$, and $1\%$ and $5\%$ tolerances for the maximum and minimum conductances, respectively. 

The effect of the variation in the parameter $\beta$ is considered inside the variations of $\alpha$. $\beta$ is modeled as a lognormal variable to have a monotonic increasing or decreasing conductance update.
Thus, the second term of the conductance update has log Gaussian variable, which is $e^z$, multiplied by $e^{\alpha n}$ where $z$ and $\alpha$ are Gaussian variables.
Since the sum of two Gaussian random variables is a Gaussian random variable, the variation of $\beta$ and $\alpha$ can be included in either one of them. 
\subsubsection{RRAM Updates for Training}\label{sec:rram_updates}
In \cite{fouda2018independent}, we proposed a method to have the resistive devices behave exactly like the learning rule where the change in each weight must be proportional to the change in the RRAM's conductance, $\Delta \mathbf{G} \propto \Delta \mathbf{w}$. 
To achieve this, the asymmetric nonlinear behavior of potentiation and depression are included in the learning rule. 
We first calculate the change in the weights for both potentiation and depression cases taking into effect the asymmetric nonlinearity of the RRAM model.
In general, the change in the LTP's and LTD's conductance due to applying $\Delta n$ is
\begin{equation}
\begin{split}
\Delta G_{LTP}=\left(G_{max}-\!G\right)\left(1-e^{-\alpha_P \Delta n}\right),\;&\text{and}\\
\Delta G_{LTD}=(G-G_{min})(e^{-\alpha_D \Delta n}-1),&
\end{split}
\end{equation}
respectively where $G$ is the previous conductance. Clearly, the relation between the rate of change in conductance and $\Delta n$ is an injective function. Thus, the number of pulses to cause $\Delta G_{LTP}$ and  $\Delta G_{LTD}$ are 
\begin{eqnarray}
\Delta n_{LTP}=-\frac{1}{\alpha_P}\ln{\left(1-\frac{\Delta G_{LTP}}{G_{max}-G(n)}\right)},\,\, \text{and} \label{DnLTP}\\
\Delta n_{LTD}=-\frac{1}{\alpha_D}\ln{\left(\frac{\Delta G_{LTD}}{G(n)-G_{min}}+1\right)}, 
\label{DnLTD}
\end{eqnarray}
\noindent respectively. 
After learning, $\Delta \mathbf{G}$ goes to $\mathbf{0}$.
As a result, $\Delta \mathbf{n}$ goes to zero as well. The update equations ((\ref{DnLTP}) and (\ref{DnLTD})) require the knowledge of the weight value, meaning a read operation is needed to calculate the required number of pulses to update. In addition, they are nonlinear functions which are expensive to implement in hardware (e.g. they require log amplifiers). 
Thus, both can be linearized as $\ln(1-x)\approx -x(1+0.5x)\approx -x$ and $\ln(1+x)\approx x(1-0.5x)\approx x$ for $x\ll 1$. 
This leads to that the potentiation and depression updates are given by 
\begin{eqnarray}
\Delta n_{LTP}=\frac{1}{\alpha_P}\frac{\Delta G_{LTP}}{G_{max}-G(n)},\quad \Delta G_{LTP}\ll G_{max}-G(n), \; \text{and} \label{DnLTP}\\
\Delta n_{LTD}=\frac{1}{\alpha_D}\frac{|\Delta G_{LTD}|}{G(n)-G_{min}}, \quad  |\Delta G_{LTD}|\ll G(n)-G_{min}, \quad \;
\label{DnLTD2}
\end{eqnarray}
It is worth mentioning that deploying the linearized update equation might lead to increasing the training time. Thus, there is a trade-off between the training time and the complexity of the weight update hardware.

\section{Synaptic Plasticity and Learning in \acp{SNN}}\label{sec:learning}
As RRAM arrays provide a scalable physical substrate for implementing neural computations and plasticity, we now turn to the modeling of synaptic plasticity. 
Synaptic plasticity in the brain is realized using some constraints as in RRAMs. 
One of these constraints is that information necessary for performing efficient weight updates must available at the physical location where the weights updates are computed and stored.

The brain is capable of learning and adapting at different time scales. Generally, learning processes operate in the hours to years range, which is thought to be implemented by synaptic plasticity and structural plasticity. 
Focusing on the former, a common synaptic plasticity process dictates that synaptic weights changes according to a modulated Hebbian-like process \cite{Gerstner_Kistler02_spikneur}, which can be written in a functional form as:
\[
\Delta W_{ij} = f(W_{ij}, S_i, S_j, M_i)
\]
where $M_i$ is some modulating function that is not yet specified. A common, biologically inspired model is \acf{STDP}.
The classical \ac{STDP} rule modifies the synaptic strengths of connected pre- and post-synaptic neurons based on the spike history in the following way:
if a post--synaptic neuron generates action potential within a time interval after the pre-synaptic neuron has fired multiple spikes then the synaptic strength between these two neurons becomes stronger (causal updates, long-term potentiation--LTP). Note that \ac{STDP} does not use the modulation term.
Formally:
\[
\Delta W_{ij}^{STDP} \propto  S_i(t) (\epsilon_{pre} \ast S_j(t)) - S_j(t) (\epsilon_{post} \ast S_i(t)) 
\]
where $\epsilon_{post}$ and $\epsilon_{pre}$ are two kernels, generally of first or second order (exponential or double exponential) filters as they relate to the neuron dynamics \refeq{eq:ulif_basic} and \refeq{eq:lif_basic}.
The convolution terms $\epsilon \ast S(t) = \int \mathrm{d}s \epsilon(s) S(t-s) $ capture the trace of the spiking activity, and serve as key building blocks for synaptic plasticity. 
These terms are key for learning in \acp{SNN} as they provide eligibility traces or memory of the neural activity history.
These traces emerge from the gradients on the neural membrane potential dynamics \cite{Zenke_Ganguli17_supesupe}.

\ac{STDP} captures the change in the postsynaptic potential amplitude in an experimental setting \cite{Bi_Poo98_synamodi} where the pair of neurons is elicited to fire at precise times. As such, it only captures a particular temporal aspect of the synaptic plasticity dynamics.
Experimental work argues that \ac{STDP} alone does not account for several observations in synaptic plasticity \cite{Shouval_etal10_spiktimi}.
Theoretical work suggested that synapses require complex internal dynamics on different timescales to achieve extensive memory capacity \cite{Lahiri_Ganguli13_memofron}.
Furthermore, error-driven learning rules derived from first principles are not directly compatible with pair-wise \ac{STDP} \cite{Pfister_etal06_optispik}.
These observations are not in contradiction with seminal work of \cite{Bi_Poo98_synamodi}, as considerable variation in LTP and LTD is indeed observed. Instead, \cite{Pfister_etal06_optispik} suggests that \ac{STDP} is an incomplete description of synaptic plasticity.

On the flip-side, normative approaches derive synaptic plasticity dynamics from mathematical principles. 
While several normative approaches exist, in the following we focus on three-factor rules that are particularly well-suited for neuromorphic applications.

\subsection{Gradient-based Learning in \acl{SNN} and Three-Factor Rules}
Three-factor rules can be viewed as extensions of Hebbian learning and \ac{STDP}, and are derived from a normative approach \cite{Urbanczik_Senn14_learby}.
The first two factors are functions of pre-synaptic activity and post-synaptic activity, and the third factor is a modulating function that is relevant to the learning task. 
Such rules have been shown to be compatible with a wide number of unsupervised, supervised, and reinforcement learning paradigms \cite{Urbanczik_Senn14_learby}, and implementations can have scaling properties comparable to that of \ac{STDP} \cite{Detorakis_etal18_neursyna}.

Three-factor rules can be derived from gradient descent on the spiking neural network \cite{Pfister_etal06_optispik,Neftci_etal19_surrgrad}. 
Such rules are often ``local'' in the sense that all the information necessary for computing the gradient is available at the post-synaptic neuron \cite{Neftci18_datapowe}.
Recent digital implementations of learning use three-factor rules, where the third factor is a modulation term that depends on an internal synaptic state \cite{Davies_etal18_loihneur} or postsynaptic neuron state \cite{Detorakis_etal18_neursyna}.

Three-factor rules are motivated by biology, where additional extrinsic factors that modulate the learning, for example, Dopamine, Acetylcholine, or Noradrenaline in reward-based learning \cite{Schultz02_gettform}, or GABA neuromodulator controlling \acl{STDP} \cite{paille2019gaba}. 
The three-factor learning rule can be written as follows: 
\begin{equation}\label{eq:3f_rule}
  \Delta W_{ij}^{3F} \propto f_{pre}(S_j(t)) f_{post}(S_i(t)) M_i
\end{equation}
where $f_{pre}$ and $f_{post}$ correspond to functions over presynaptic and post synaptic variables, respectively, and $M_i$ is the modulating term of postsynaptic neuron $i$.
The modulating term is a task-dependent function, which can represent error, surprise, or reward.  

The equivalence of \acp{SNN} with artificial neural networks discussed in \refsec{sec:lif} paired with synaptic plasticity derived from gradient descent suggests that the same approaches used for training artificial networks can be applied to \acp{SNN}.
In other words, the synaptic plasticity rule can be formulated in a way that it optimizes a task-relevant objective function \cite{Neftci18_datapowe}.
A machine learning description of \ac{SNN} training consists of three parts: The objective function, the (neural network) model and the optimization strategy.
The objective, noted $\mathcal{L}(\mathbf{S}(\Omega),\mathbf{S_{data}})$ is a scalar function describing the performance of the task at hand (e.g. classification error, reconstruction error, free energy, etc.), where $\Omega$ are trainable parameters and $\mathbf{S}$,$\mathbf{S_{data}}$ are neural states (spikes, membrane potentials, etc.) and input spikes, respectively dictated by the \ac{SNN} dynamics. 
If operating in a firing rate mode (where spike counts or mean firing rates are carriers of task-level information),  $\mathbf{S}$, and $\mathbf{S_{data}}$ can be interpreted as firing rates instead.
The optimization strategy consists of a parameter update derived from gradient descent on $\mathcal{L}$.
If this update rule can be expressed in terms of variables that are local to the connection, then the learning rule will be called a synaptic plasticity rule.

Gradient-based approaches have been used in a wide range of work. For examples, the Tempotron is a learning rule using a membrane potential-based objective function to learn to discriminate between inputs on the basis of the spike train statistics \cite{Gutig_Sompolinsky06_tempneur}; \cite{Pfister_etal06_optispik} expresses the neuron model as a stochastic generative model and derive learning rules by maximizing the likelihood of a target spike train. Gradient-based approaches identify (non-unique) relationships between the \ac{STDP} parameters and those of the neural and synaptic dynamics; and SpikeProp \cite{Bohte_etal00_spikback}, a spike-based gradient backpropagation algorithm. Several other approaches that can collectively be described as surrogate gradient descent \cite{Neftci_etal19_surrgrad} rely on approximations of the gradient to perform \ac{SNN} parameter updates \cite{Huh_Sejnowski17_graddesc,Shrestha_Orchard18_slayspik,Anwani_Rajendran15_normappr,Zenke_Ganguli17_supesupe}.

While the above models are computationally promising, there are important challenges in learning multilayer models on a physical substrate such as the brain: The physical substrate defines which variables are available to which processes and when. 
This is in stark contrast to von Neumann computers where learning processes have access to shared memory. One consequence of this limitation is the weight transport problem of gradient backpropagation, where the symmetric transpose of the weights is necessary to train deep networks. In many cases, however, the neurons and weight tables cannot be "reversed" in this fashion. Another less studied consequence is the temporal locality due to the continual dynamics of \acp{SNN}: solving the credit assignment problem in recurrent neural networks requires some memory of the previous states and inputs, either in the form of buffers or eligibility traces \cite{Williams_Zipser89_learalgo}. 
Both of these problems, namely the weight transport problem and temporal credit assignment, must be solved in order to successfully implement memristor-based machine inference and learning.
Below, we describe some promising approaches that overcome these two problems. 

\begin{figure*}
    \begin{center}
      \includegraphics[width=.8\textwidth] {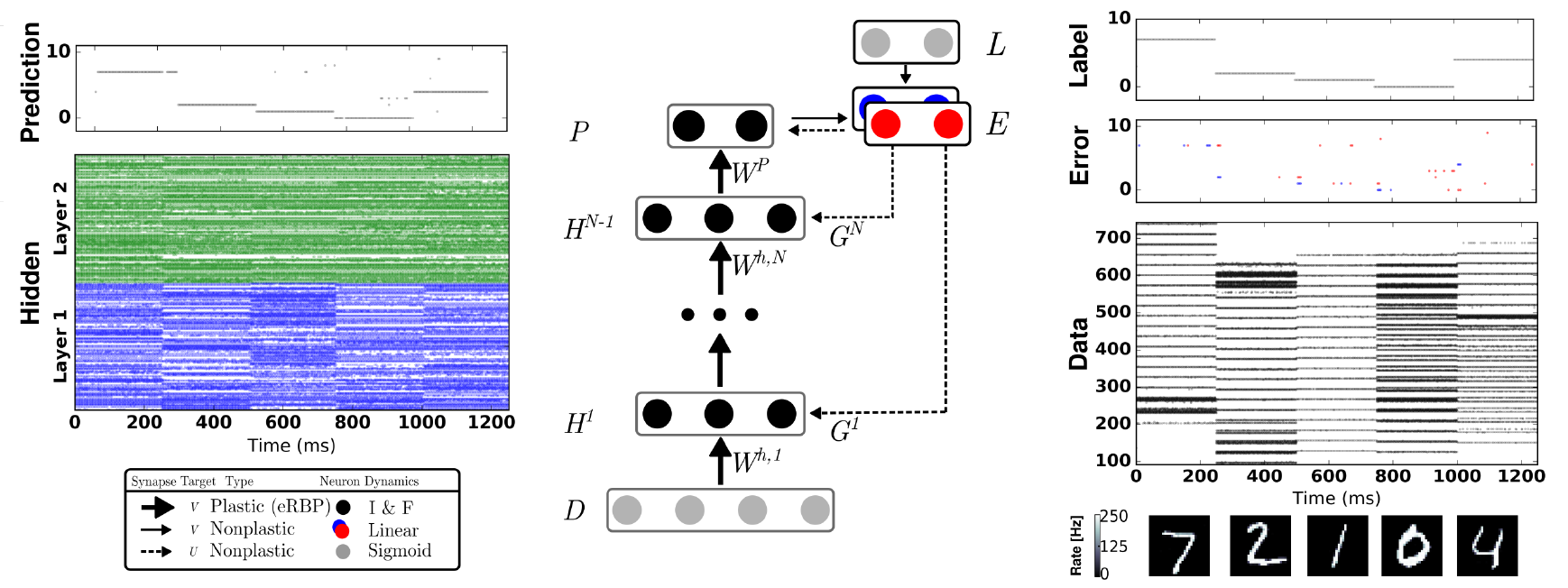}
      \hspace{3pt}
    \end{center}
    \caption{
      \label{fig:erbp} 
      \emph{Network Architecture for Event-driven Random Backpropagation (eRBP)}.
    The network consists of feed-forward layers (784-200-200-10) for prediction and feedback layers for supervised training with labels (targets) $\mathcal{L}$.
    Full arrows indicate synaptic connections, thick full arrows indicate plastic synapses, and dashed arrows indicate synaptic plasticity modulation.
    Neurons in the network indicated by black circles were implemented as two-compartment \ac{LIF} neurons.
    The top-layer errors are proportional to the difference between labels (L) and predictions (P) and is implemented using a pair of neurons coding for positive error (blue) and negative error (red).
    Each hidden neuron receives inputs from a random combination of the pair of error neurons.
    Output neurons receive inputs from the pair of error neurons in a one-to-one fashion. Reproduced from \cite{Neftci_etal17_evenrand}}.
\end{figure*}

\paragraph{Feedback Alignment and Event-driven RBP}
One way to alleviate the weight transport problem is to replace the weights used in backpropagation with fixed, random ones \cite{Lillicrap_etal16_randsyna}.
Theoretical work suggests that the network adjusts its feed-forward weights such that they align with the (random) feedback weights, which is arguably equally good in communicating gradients.
This approach is naturally extended to \acp{SNN} to overcome the weight transport problem. Event-driven Random Back Propagation (eRBP) is one such rule that extends feedback alignment to meet the other constraints of learning in \acp{SNN}, namely that weight updates are event-based (no separate forward and backward phases) and errors are maintained on a dedicated compartment of the neuron, rather than in a globally accessible buffer. Because it is local, it can be implemented as a presynaptic spike-driven plasticity rule modulated by top-down errors and gated by the state of the postsynaptic neuron and is simple to implement. The learning rule can be written as:
\begin{equation}\label{eq:erbp_rule}
\begin{split}
M_i &= \sum_k g_{ik} Error_k\\
\Delta W_{ij}^{eRBP} &\propto  M_i Boxcar(U_i) S_j(t)
\end{split}
\end{equation}
where $g_{ik}$ are fixed, random weights and $Boxcar$ is a symmetric function that is equal to 1 in the vicinity of $U_i=0$, and zero otherwise. 
Here, $M_i$ represents the state of the neural compartment that modulates the plasticity rule according to top-down errors. 
Its functionality is to maintain a trace of $Error_k = target_k - S_k$ when an input spike occurs.
ERBP solves the nonlocality problems, leading to remarkably good performance on MNIST handwritten digit recognition tasks (\reffig{fig:erbp}), achieving close to $2\%$ error compared to $1.9\%$ error using gradient backpropagation on the same network architecture.

One limitation \ac{eRBP} is related to the ``loop duration'', \emph{i.e.} the duration necessary from the input onset to a stable response in the error neurons. 
A related problem is layerwise locking in deep neural networks: because errors are backpropagated from the top layers, hidden layers must wait until the prediction is made available \cite{Jaderberg_etal16_deconeur}. 
This duration scales with the number of layers, limiting \ac{eRBP} scalability for very deep networks.
The loop duration problem is caused by the temporal dynamics of the spiking neuron, which are not taken into account in \refeq{eq:erbp_rule}. 

This problem can be partly overcome by maintaining traces of the input spiking activity, and a solution was reported in \cite{Zenke_Ganguli17_supesupe}. 
Their rule, called Superspike is derived from gradient descent over the \ac{LIF} neuron dynamics, resulting in the following three-factor rule:
\begin{align}\label{eq:ss_rule}
\Delta W_{ij}^{SS} \propto  \alpha \ast (Error_i \rho'(U_i) (\epsilon_{pre} \ast S_j)))
\end{align}
where $\rho$ describes the slope of the activation function $\rho$ at the membrane potential $U_i$, and $\epsilon_{pre}$ here is the response of the post-synaptic neuron to a pre-synaptic spike (the impulse response function at $U$).

Interestingly, both \refeq{eq:ss_rule} and \refeq{eq:erbp_rule} rules are reminiscent of \ac{STDP} but include further terms that vary according to some external modulation, itself related to some task. 
Unsurprisingly, the three terms in \refeq{eq:3f_rule} can be related to the classical Widrow-Hoff (Delta) rule, where the first term is the error, the second is the derivative of the output activation function, and the third term is the input.

The loop duration is only partly solved with \refeq{eq:ss_rule}, as $\alpha$ and $\epsilon$ introduce memory of the previous activity into the synapses. However, this is only an approximation, as the dynamics of every layer leading to the top layer must be taken into account with one additional temporal convolution per layer. As a result, \refeq{eq:ss_rule} and \refeq{eq:erbp_rule} do not scale well with multiple layers. 

\paragraph{Local Errors}
A more effective method to overcome the loop duration and the layerwise locking problem is to use synthetic gradients: gradients that can be computed locally, at every layer. 
Synthetic gradients were initially proposed to decouple one or more layers from the rest of the network to prevent layerwise locking \cite{Jaderberg_etal16_deconeur}.
Synthetic gradients usually involve an outer loop consisting of a full backpropagation through the network. 
While this provides convergence guarantees, a full backpropagation step cannot be done locally in spiking neural networks. 
Instead, relying on initialization of the local random classifier weights and forgoing the outer loop training yields good empirical results \cite{Mostafa_etal17_deepsupe}.

\begin{figure}
\begin{center}
  \includegraphics[height=.32\textheight]{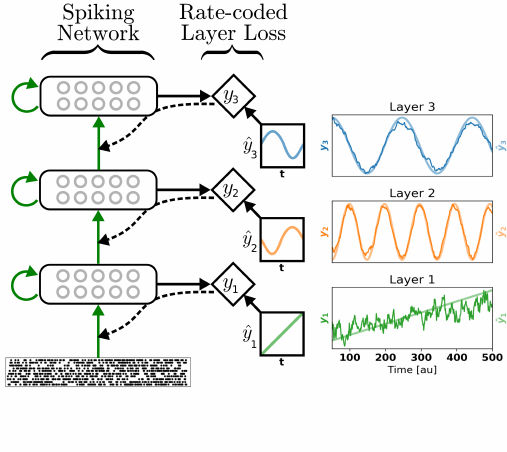}
  \hspace{.05\textwidth}
  \includegraphics[height=.32\textheight]{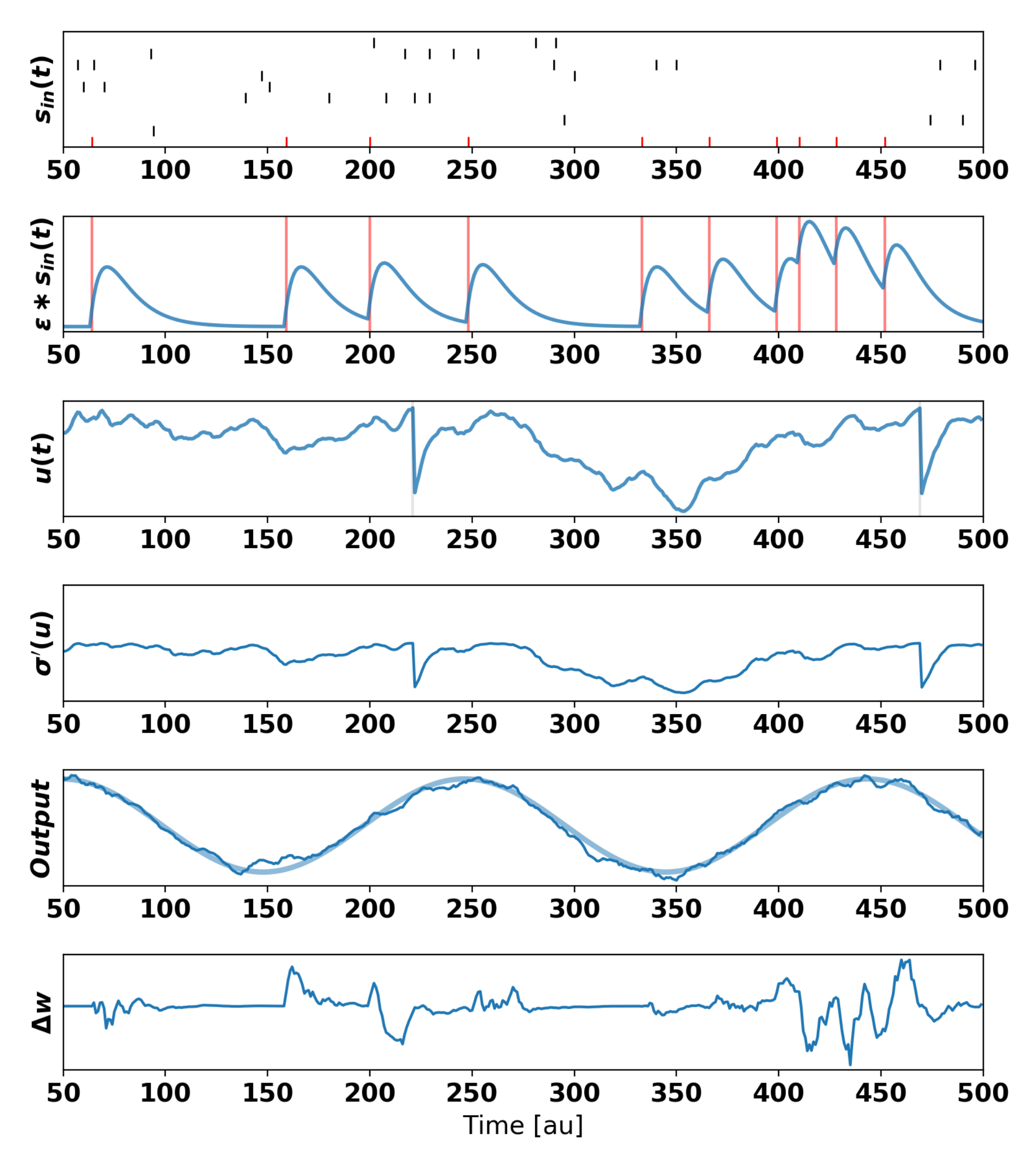}
\end{center}
  \caption{\label{fig:dcll_illustration}\emph{Deep continuous local learning example}. (Left) Each layer consists of spiking neurons with continuous dynamics. Each layer feeds into a local classifier through fixed, random connections (diamond-shaped, $y$). The classifier is trained to produce auxiliary targets $\hat{y}$. Errors in the local classifiers are propagated through the random connections to train the input weights, but no further (curvy, dashed line). To simplify the learning rule and enable linear scaling of the computations, the cost function is formulated using a rate code. The states of the spiking neurons (membrane potential, synaptic states, refractory state) are carried forward in time. Consequently, even in the absence of recurrent connections, the neurons are stateful in the sense of recurrent neural networks.
  (Right) Snapshot of the neural states illustrating the DECOLLE learning rule in the top layer. In this example, the network is trained to produce three time-varying pseudotargets $\hat{y}_1$, $\hat{y}_2$ and $\hat{y}_3$. Reproduced from \cite{Kaiser_etal18_synaplas}}.
\end{figure}

Using layerwise local classifiers \cite{Mostafa_etal17_deepsupe}, the gradients are computed locally using pseudo-targets (for classification, the labels themselves). 
To take the temporal dynamics of the neurons into account, the learning rule is similar to SuperSpike \cite{Zenke_Ganguli17_supesupe}.
However, the gradients are computed locally through a fixed random projection of the network activities into a local classifier. 
Learning is achieved using a local rate-based cost function reminiscent of readout neurons in liquid state machines \cite{Maass_etal02_realcomp}, but where the readout is performed over a fixed random combination of the neuron outputs.
The readout does not have a temporal convolution term in the cost function, the absence of which enables linear scaling, and does not prevent learning precise temporal spike trains (\reffig{fig:dcll_illustration}).
The resulting learning dynamics are called DEep COntinuous Local LEarning (DECOLLE), and written:
\begin{align}\label{eq:dcll_rule}
  \Delta W_{ij}^{DECOLLE} \propto  (\sum_k g_{ik} Error_k) \rho'(U_i) (\epsilon_{pre} \ast S_j)).
\end{align}
Here, the error is computed with respect to a random linear combination of the neuron outputs: $Error_k = target_k - \sum_i g_{ik}S_i $.
While SuperSpike scales at least quadratically with the number of neurons, learning with local errors scales linearly \cite{Kaiser_etal18_synaplas}. 
Linearity greatly improves the memory and computational cost of computing the weight updates and simplifies potential RRAM implementations (see \refsec{sec:3f_rram}).

\paragraph{Independent Component Analysis with Three-Factor Rule}
Independent component analysis (ICA) is a very powerful tool to solve the cocktail party problem (blind source separation), feature extraction (sparse coding) and can be utilized in many applications such as de-noising images, Electroencephalograms (EEG) signals, and telecommunications \cite{hyvarinen2004independent}. ICA consists of finding mutually independent and non-Gaussian hidden factors (components), $\mathbf{s}$, that form a set of signals or measurements, $\mathbf{x}$. This problem can be mathematically described for linearly mixed components as follows  
$\mathbf{x}=\mathbf{A}\mathbf{s}$ where $A$ is the mixing matrix. Both $\mathbf{A}$ and $\mathbf{s}$ are unknowns. {In order to find the independent components (sources), the problem can be formulated as $\mathbf{u}=\mathbf{W}\mathbf{x}$ where $\mathbf{x}$ is the mixed signals (inputs of ICA algorithm), $\mathbf{W}$ is the weight matrix (demixing matrix), $\mathbf{u}$ is the outputs of the ICA algorithm (independent components).} ICA's strength lies in  utilizing the mutual statistical independence of components to separate the sources. 

Recently, Isomura and Toyoizumi have proposed a biological plausible learning rule called \textit{Error-Gated Hebbian Rule (EGHR)} inspired from the standard Hebbian rule \cite{isomura2016local} that enables local and efficient learning to find the independent components. The EGHR learning rule can be written as 
\begin{equation}
\mathbf{\Delta{W}}^{EHGR}=\eta \langle\left(E_o-E(\mathbf{u})\right)g(\mathbf{u})\mathbf{x}^T\rangle
\end{equation}
\noindent where $\eta$ is the learning rate, $\langle\cdot\rangle$ is the expectation over the ensemble (training samples), $g(u_i)x_j$ is the Hebbian learning rule, $g(u_i)$, $x_j$ are the postsynaptic and presynaptic terms of the neuron, respectively, $(E_o-E(\mathbf{u}))$ is the global error signal which consists of $E_o$ which is a constant, and $E(\mathbf{u})$ which is the surprise or reward that guides the learning. The cost function of EGHR is defined as $\mathcal{L}=c\frac12\langle(E_0-E(\mathbf{u}))^2\rangle$. It was proven mathematically and numerically that this learning rule is robust, stable and its equilibrium point is proportional to the inverse of the mixture matrix, \emph{i.e.} the solution of ICA. This learning rule is a clear example of three-factor learning where the modulating is represented in the surprise,  $\left(E_o-E(\mathbf{u})\right)$. 
This learning rule is a three-factor rule and can be performed using spiking neuron by following an implementation similar to  \cite{savin2010independent}.

ICA assumes that the sources are linearly mixed using a mixture matrix $\mathbf{A}$. The final weight matrix, $\mathbf{w}$, should converge to $c \mathbf{A}^{-1}$ which is still a valid solution since $c$ is a scaling factor. 
As previously discussed, $\Delta G_{ij}$ can be replaced by $\eta' \Delta w_{ij}$ to match the synaptic dynamics \cite{fouda2018independent}, where $\eta'$ is the scaled learning rate, and $\Delta w_{ij}$ is given by EGHR. Thus, the final equations for potentiation and depression pulses can be written as follows: 
\begin{eqnarray}
\Delta n_{ij}|_{LTP}\approx \frac{1}{\alpha_P}\left(\frac{\eta' (E_o-E(\mathbf{u}))g(u_j)x_i}{G_{max}-G_{ij}(n)}\right),\,\, \text{and} \\
\Delta n_{ij}|_{LTD}=-\frac{1}{\alpha_D}\left(\frac{\eta' (E_o-E(\mathbf{u}))g(u_j)x_i}{G_{ij}(n)-G_{min}}\right), 
\end{eqnarray}
\noindent respectively. By programming the RRAMs using the previous equations, the circuit behaves as required and compensates for the asymmetric nonlinearity of the devices.

\begin{figure}[!t]
\centering
\vspace{-0.2in}
\subfloat[]{\includegraphics[width=0.45\linewidth,]{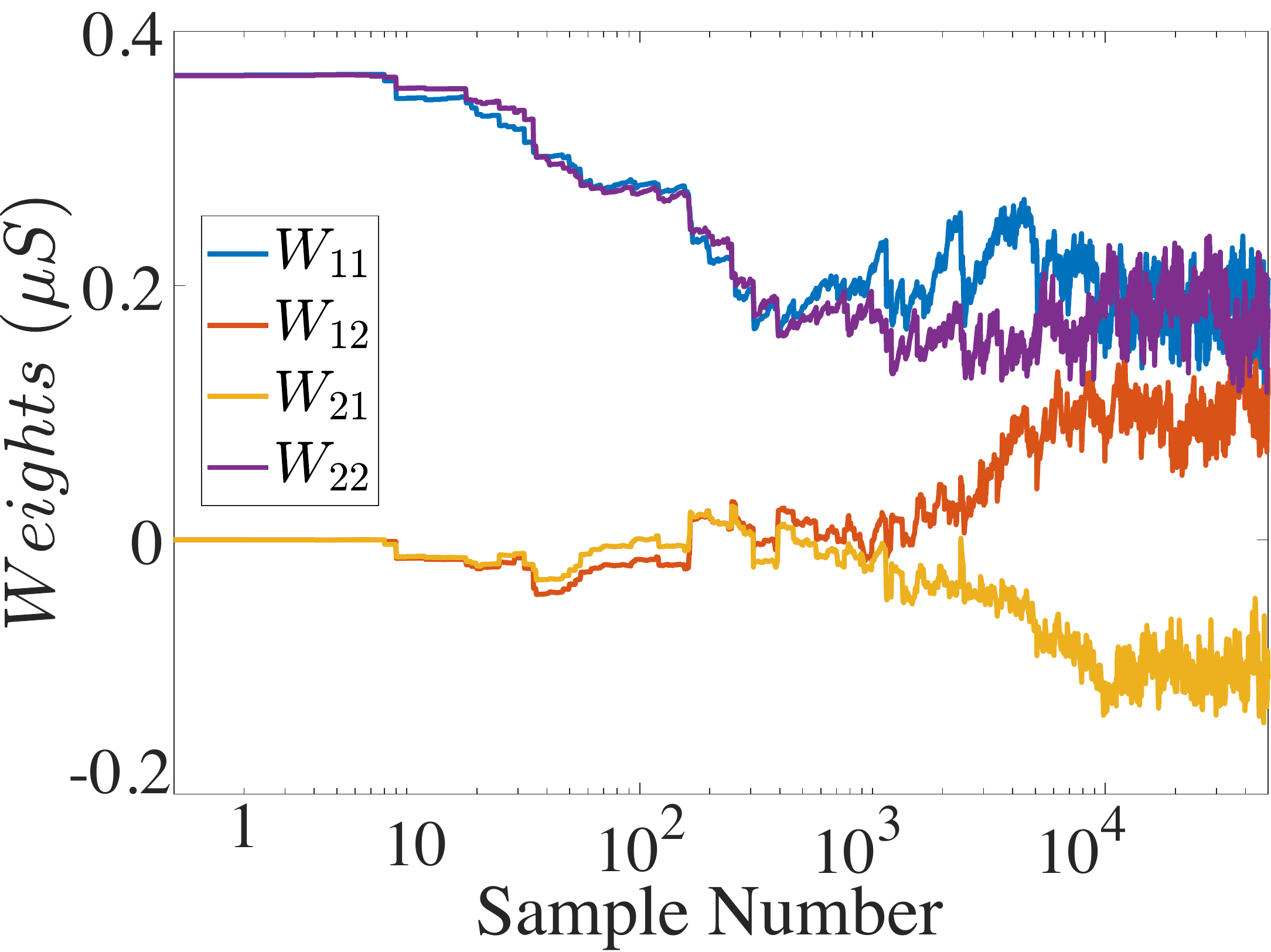}%
\label{weights}}
\hfil
\vspace{-0.01in}
\subfloat[]{\includegraphics[width=0.5\linewidth]
{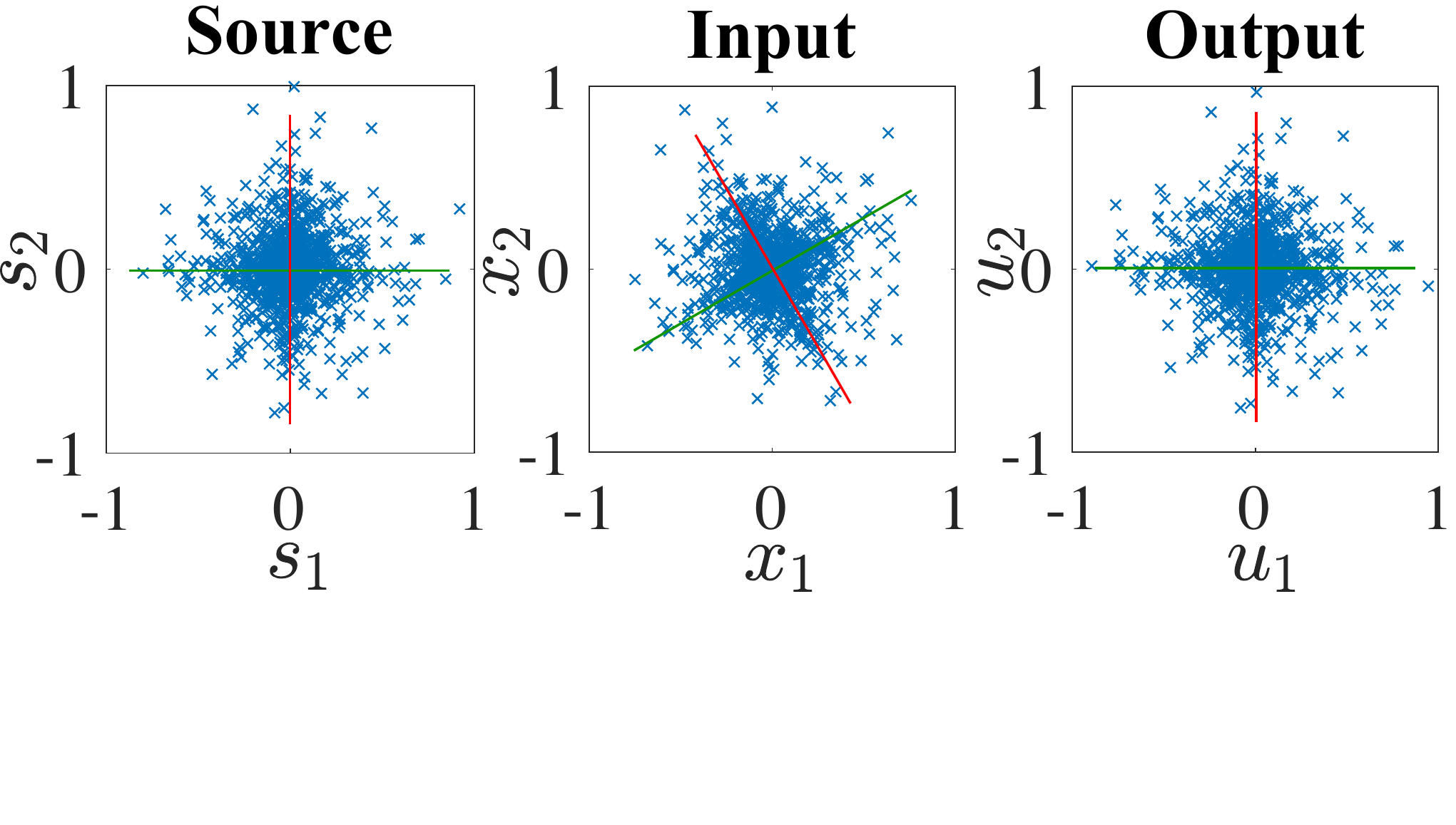}%
\label{Results}}
\vspace{-0.05in}
\caption{The online training results versus training time (a) Evolution of the weights, and (b) Visual results of the input and the output. Reproduced from \cite{fouda2018independent}}
\label{FigResults}
\end{figure}

As a test bench for the proposed technique, we considered two Laplacian random variables that are generated and mixed using a mixture matrix which is set to a rotation matrix $\mathbf{A}=(\cos\theta, -\sin\theta; \sin\theta, \cos\theta)$ with $\theta=\pi/6$. Figure \ref{FigResults} shows the results of the online learning of independent components of the mixed signals. 
Figure \ref{weights} shows the weight evolution during the training. Clearly, there are some oscillations in the weights around the final solution after $10^4$ samples because of the continuous on-line learning and the device variations. This can be avoided by using dynamic learning rate such as Adaptive Moment Estimation (Adam). A visual representation of the signals before mixing, after mixing and after training is shown in Fig. \ref{Results} which depicts the similarity between the source and output signals.

\section{Stochastic Spiking Neural Networks}
Up to now, we have considered fully deterministic neuron and synapse models. However, as discussed in \refsec{sec:rram_updates}, the writing (and in some cases, reading) of RRAMs values are stochastic. 
Additionally, analog VLSI neuron circuits have significant variability across neurons due to fabrication mismatch (fixed pattern noise) and behave stochastically due to noise intrinsic to the device operation.
The variability at the device level can be taken in to account in \ac{SNN} models and sometimes be exploited for improving learning performance and implementing probabilistic inference \cite{Querlioz_etal15_bioiprog,Naous_etal16_memrneura}. Here, we list avenues for implementing online learning in memory devices that exploit the stochasticity in the neurons and synapses.

A stochastic model of the neurons can be expressed as:
\begin{equation}\label{eq:srm_neuron}
  P( S_i | \mathbf{s} ) = \rho(U_i(t)) 
\end{equation}
where $\rho_i$ is the stochastic intensity (the equivalent of the activation function in artificial neurons), and $\eta$ and $\epsilon$ are kernels that reflect neural and synaptic dynamics, e.g. refractoriness, reset and postsynaptic potentials \cite{Gerstner_Kistler02_spikneur}.
The stochastic intensity can be derived or estimated experimentally if the noiseless membrane potential ($U_i(t)$) can be measured at the times of the spike \cite{Jolivet_etal06_predspik}.
This type of stochastic neuron model drives numerous investigations in theoretical neuroscience and forms the starting point for other types of adapting spiking neural networks capable of efficient communication \cite{Zambrano_Bohte16_fasteffi}.

\subsection{Learning in Stochastic Spiking Neural Networks}
Neural and synaptic unreliability can induce the necessary stochasticity without requiring a dedicated source of stochastic inputs, for example, the unreliable transmission of synaptic vesicles in biological neurons. This is a well-studied phenomenon \cite{Katz66_nervmusc,Branco_Staras09_probneur}, and many studies suggested it as a major source of stochasticity in the brain \cite{Faisal_etal08_noisnerv,Abbott_Regehr04_synacomp,Yarom_Hounsgaard11_voltfluc,Moreno-Bote14_poisspik}. 
In the cortex, synaptic failures were argued to reduce energy consumption while maintaining the computational information transmitted by the post-synaptic neuron \cite{Levy_Baxter02_enerneur}.
More recent work suggested synaptic sampling as a mechanism for representing uncertainty in the brain, and its role in synaptic plasticity and rewiring \cite{Kappel_etal15_netwplas}.

Strikingly, the simplest model of synaptic unreliability, a ``\emph{blank-out}'' synapse, can improve the performance of spiking neural networks in practical machine learning tasks over existing solutions, while being extremely easy to implement in hardware \cite{Goldberg_etal01_probsyna}, and often naturally occurring in emerging memory technologies \cite{Saghi_etal15_plasmemr,Al-Shedivat_etal15_inhestoc,Yu_etal13_stoclear}.

One approach to learning with such neurons and synapses is Event-Driven Contrastive Divergence (ECD), using ideas borrowed from Contrastive Divergence in restricted Boltzmann machines \cite{Hinton02_traiprod}. 
The stochastic neural network produces samples from a probability distribution, and \ac{STDP} carries out the weight updates according to the Contrastive Divergence rule in an online, asynchronous fashion. 
In terms of the three-factor rule above, ECD can be written:
\begin{equation}
\Delta W^{ECD}_{ij} = M_i(t) \Delta W^{STDP}_{ij}
\end{equation}
where $M_i(t) = 1$ during the ``data'' phase and $M_i(t)=-1$ during the ``reconstruction'' phase.
These neural networks can be viewed as a stochastic counterpart of Hopfield networks \cite{Hopfield82_neurnetw}, but where stochasticity is caused by multiplicative noise at the connections (synapses) or at the nodes (neurons).
\begin{figure*}
    \centering
    \includegraphics[width=.9\textwidth]{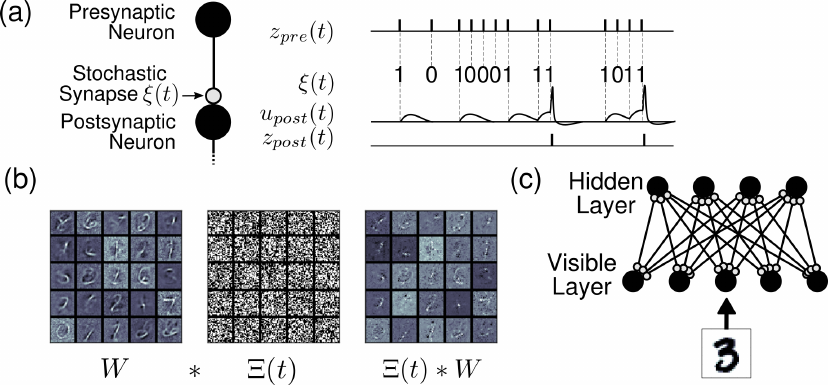}
    \caption{\label{fig:srbm} The \acf{SSM}. (a) At every occurrence of a pre-synaptic event, a pre-synaptic event is propagated to the post-synaptic neuron with probability $p$. (b) Synaptic stochasticity can be viewed as a continuous DropConnect method \cite{Wan_etal13_reguneur} where weights are masked by a binary matrix $\Theta(t)$, where $\ast$ denotes element-wise multiplication. (c) \ac{SSM} Network architecture, consisting of a visible and a hidden layer. Reproduced from \cite{Neftci_etal16_stocsyna}}.
\end{figure*}

ECD requires symmetric weights, which is difficult to achieve due to the weight transport problem discussed above. 
A variation of ECD called random Contrastive Hebbian learning (rCHL) \cite{Detorakis_etal18_conthebb}, replaces the transpose of the synaptic weights with fixed random matrices. 
This was performed similarly to Feedback Alignment (FDA)~\cite{Lillicrap_etal16_randsyna}. 
Contrastive Hebbian Learning (CHL) is similar to Contrastive Divergence, but it employs continuous nonlinear dynamics at the neuronal level.
Like Contrastive Divergence, it does not rely on a special circuitry to compute gradients (but can be interpreted as the gradient of an objective function), allows information to flow in a coupled, synchronous way, and is grounded upon Hebb's learning rule. 
CHL uses feedback to transmit information from the output layer to hidden(s) layer(s), and in instances when the feedback gain is small (such as in the clamped phase), has been demonstrated by Xie and Seung to be equivalent to Backpropagation~\cite{Xie_Seung03_equiback}.
Using this approach, the information necessary for learning propagates backward, though it is not transmitted through the same axons (as required in the symmetric case), but instead via separate pathways or neural populations.

Equilibrium Propagation (EP) describes another modification of CHL that generalizes the objective functions it can solve and improve on its theoretical groundings.
In EP the neuron dynamics are derived from the energy function, however, EP requires symmetric weights. 
The energy function used in Equilibrium Propagation includes a mean-squared error term on the output units, allowing the output to be weakly clamped to the true outputs. (e.g. labels). 
The neuron model takes a form which is reminiscent of the \ac{LIF} neuron. The recurrent dynamics in the network affect the other layers in the network, similarly to CHL. 
Both rCHL and EP were formulated for continuous (rate-based) neurons, although their implementation with spikes is straightforward following the same approach as \acp{SSM}.

Learning in stochastic neuron networks can also be performed using the surrogate gradient approach and three-factor rules. 
In this case, a simple approach is to use the stochastic intensity $\rho$ as a drop-in replacement of the neural activation function for purposes of computing the weight updates \cite{Neftci_etal19_surrgrad}.
In this case, stochasticity plays a regularization role similar to dropout in deep learning \cite{Neftci_etal17_evenranda} and weight normalization \cite{Neftci17_stocsyna} an online modification of batch normalization.

\subsection{Three Factor Learning in Memristor Arrays}\label{sec:3f_rram}
So far, we have discussed how to implement gradient-based learning as local synaptic plasticity rules in \acp{SNN}. 
In many cases, gradient-based learning provides superior results compared to \ac{STDP} and take the form of three-factor rules.
These rules are biologically credible since pre- and post-synaptic activities are available at the level of the neuron and neurotransmitters in the brain can carry the extrinsic factor.
However, besides the LTP and LTD asymmetry problems already discussed, the implementation of the three-factors in memristor arrays come with certain challenges. 
The first challenge concerns the implementation of the synaptic traces.
In certain simple cases, such as when the subthreshold neural dynamics are linear, only one trace for each neuron involved in the learning connections is required for learning \cite{Kaiser_etal18_synaplas}, similar to the \ac{STDP} case.
Previous work has demonstrated \ac{STDP} learning in RRAM and hence capture some form of a neural trace.
The majority of these include additional CMOS circuitry in a ``hybrid configuration'' \cite{Ielmini18_braicomp} to enable \ac{STDP}.
The simplest of these implementations consists of a 2T1R configuration that enables an update when both terminals are high (both spike).
While this is sufficient for the case where $\alpha=\beta=0$, an additional mechanism that filters the spike is necessary to recover STDP like curves when $\alpha>0$ or $\beta>0$.
This can be achieved with a circuit that is similar to that of the synapse \cite{Bartolozzi_Indiveri08_silisyna} or calcium variables \cite{Mitra_etal06_vlsispik}.
For more complex neuron dynamics (such as non-linear neuron dynamics), then at least one trace \emph{per synapse} is required, \cite{Bellec_etal19_biolinsp}, and ideally, one trace per connection and per neuron \cite{Williams_Zipser89_learalgo}.

Since the synaptic trace dynamics follow similar first order (RC) dynamics, the same circuits used to implement first-order synaptic dynamics can be used to implement synaptic traces \cite{Bartolozzi_Indiveri06_silisyna} or dedicated calcium dynamics circuits \cite{Huayaney_etal16_learsili}.
However, scalability can become an issue: the same amount of memory for storing the weights is necessary for computing the traces, but the latter must be carried out continuously.
One potential solution comes from recent work in using diffusive memristors to implement the leaky dynamics of integrate and fire neurons \cite{Wang_etal18_fullmemr}, which can be used for computing neural traces.

The second issue concerns the modulation. 
In many gradient-based three-factor rules, the modulation of the learning rule is specific to each neuron, not each synapse.
This means that a similar approach to \ac{eRBP} \refeq{eq:erbp_rule}, where a separate neuron compartment is used for maintaining the modulation factor can in principle be used in memristor arrays, \emph{i.e.} the weight update can consist in the two factors $(\epsilon_{pre} \ast S_j))$, where $M_i \rho'(U_i)$. 

Additionally, the variability in the conductance reading and writing can cause the learning to fail or slow down. Independent noise in the read or write is not a problem and can even help learning, as discussed in the stochastic \ac{SNN} section. Fixed pattern noise, however, can be problematic as it translates into variable learning rates per weight and can impair learning.

\section{Concluding Remarks}
In this chapter, we presented neural and synaptic models for learning in spiking neural networks. In particular, we focused on synaptic plasticity models that use approximate gradient-based learning that is potentially compatible with a neuromorphic implementation and memristor arrays. 
Gradient-based learning in spiking neural networks generally provide the best performances on many applications, such as image recognition, probabilistic inference, and ICA. The mathematical modeling of the synaptic plasticity revealed that these dynamics take the form of three-factor rules, which can be viewed as a type of modulated Hebbian learning. While Hebbian learning or its spiking counterpart, \ac{STDP} have been previously demonstrated in memristors, three-factor rules also require modulation of the learning at the postsynaptic neurons.
Furthermore, if the neuron and synapse model are equipped with temporal dynamics, then it may become necessary to maintain pre-synaptic and post-synaptic activity traces in the crossbar to address the temporal credit assignment problem. Through the mathematical models, we identified which approaches are viable for implementing the modulation and the neural traces with memristors.

\bibliographystyle{plainnat} 
\bibliography{biblio_unique_alt,foudaRef}

\end{document}